\documentclass[%
 aip,
 amsmath,amssymb,
 reprint,%
]{revtex4-1}

\usepackage{graphicx}
\usepackage{dcolumn}
\usepackage{bm}

\usepackage[utf8]{inputenc}
\usepackage[T1]{fontenc}
\usepackage{mathptmx}
\usepackage{etoolbox}

\usepackage{color}
\usepackage{soul}
\newcommand{\rp}[1]{\textcolor{black}{#1}}

\makeatletter
\def\@email#1#2{%
 \endgroup
 \patchcmd{\titleblock@produce}
  {\frontmatter@RRAPformat}
  {\frontmatter@RRAPformat{\produce@RRAP{*#1\href{mailto:#2}{#2}}}\frontmatter@RRAPformat}
  {}{}
}%
\makeatother
\begin{document}

\preprint{AIP/123-QED}

\title[Thermodynamics, formation dynamics and structural correlations in the bulk amorphous phase of the phase-field crystal model]{Thermodynamics, formation dynamics and structural correlations in the bulk amorphous phase of the phase-field crystal model}

\author{Shaho Abdalla}
\affiliation{Department of Mathematical Sciences, Loughborough University, Loughborough LE11 3TU, UK}
\affiliation{Interdisciplinary Centre for Mathematical Modelling,
Loughborough University, Loughborough LE11 3TU, UK}
 
\author{Andrew J. Archer}%
\homepage{https://www.lboro.ac.uk/departments/maths/staff/andrew-archer/}
\affiliation{Department of Mathematical Sciences, Loughborough University, Loughborough LE11 3TU, UK}%
\affiliation{Interdisciplinary Centre for Mathematical Modelling,
Loughborough University, Loughborough LE11 3TU, UK}

\author{L\'aszl\'o Gr\'an\'asy}
\affiliation{Wigner Research Centre for Physics, P.O. Box 49, H-1525 Budapest, Hungary}
\affiliation{BCAST, Brunel University, Uxbridge, Middlesex UB8 3PH, UK}

\author{Gyula I. T\'oth}
 \homepage{https://www.lboro.ac.uk/departments/maths/staff/gyula-toth/}
 \email{g.i.toth@lboro.ac.uk}
\affiliation{Department of Mathematical Sciences, Loughborough University, Loughborough LE11 3TU, UK}%
\affiliation{Interdisciplinary Centre for Mathematical Modelling,
Loughborough University, Loughborough LE11 3TU, UK}

\date{\today}

\begin{abstract}
We investigate bulk thermodynamic and microscopic structural properties of amorphous solids in the framework of the phase-field crystal (PFC) model. These are metastable states with a non-uniform density distribution having no long-range order. From extensive numerical simulations we determine the distribution of free energy density values in varying size amorphous systems and also the point-to-set correlation length, which is the radius of the largest volume of amorphous one can take while still having the particle arrangements within the volume determined by the particle ordering at the surface of the chosen volume. We find that in the thermodynamic limit, the free energy density of the amorphous tends to a value that has a slight dependence on the initial state from which it was formed -- i.e.\ it has a formation history dependence. The amorphous phase is observed to form on both sides of the liquid linear-stability limit, showing that the liquid to amorphous transition is first order, with an associated finite free energy barrier when the liquid is metastable. In our simulations this is demonstrated when noise in the initial density distribution is used to induce nucleation events from the metastable liquid. Depending on the strength of the initial noise, we observe a variety of nucleation pathways, in agreement with previous results for the PFC model, and which show that amorphous precursor mediated multi-step crystal nucleation can occur in colloidal systems. 
\end{abstract}

\maketitle

\section{Introduction}

Amorphous materials are widely used in electronics, energy, medical, and environmental technologies. For instance, hydrogenated amorphous silicon is used in thin-film transistors and solar cells \cite{GASPARI2018117}, while drug molecules are formulated as amorphous solids in order to enhance the solubility when they have weak water-solubility \cite{babu2011solubility}, metallic glasses form super-hard metals with a variety of applications \cite{telford2004case}, oxide glasses and partly crystalline glass-ceramics are widely used in biomedical \cite{biomed_1,biomed_2} and automotive/aerospace applications \cite{aerospace}, in optical fibers, glass fiber reinforced materials \cite{reinforced} and in phase-change-optical-recording \cite{optrec_1,optrec_2}. It is well-known that the macroscopic material properties of amorphous materials are determined by the underlying microscopic structure. Amorphous materials for advanced industrial applications are manufactured through carefully controlled solidification processes. Achieving the desired control is still challenging and time consuming, requiring detailed understanding of the material structure formation process. Thus, it is crucial to develop a good theoretical understanding of the underlying physical phenomena driving microstructure evolution such as the interplay between solid nucleation and the glass transition. These phenomena are amongst the least understood in solid state physics, so tackling them is important. Ref.~\onlinecite{berthier2011theoretical} gives a good introduction to various of the theoretical approaches used to understand amorphous materials. Here, we present an analysis of the formation dynamics, microstructure and thermodynamics of the amorphous phase formed by the so-called phase-field crystal (PFC) model. This is a model for \rp{the density profile in materials}, which is both sufficiently coarse-grained so that it can be used for mesoscopic sized systems, yet at the same time it incorporates an (atomic) particle-level description of the solid structure \cite{PFCreview2012}. \rp{Therefore}, the PFC seems an ideal model for investigating the properties of amorphous materials in relation to applications such as those mentioned above. The work presented here will underpin such future uses of the model.

We determine the conditions under which the PFC model forms amorphous states, instead of \rp{the thermodynamically stable bulk crystalline phase}. We also investigate thermodynamic properties of the amorphous \rp{states}. Additionally, we calculate the point-to-set correlation length $l_0$ \cite{berthier2011theoretical, bouchaud2004adam} in the amorphous structures formed by the PFC model and how it depends on the state point of the system. \rp{Originally developed to be measured from particle-based simulations, $l_0$ is the maximal radius of a spherical volume for which the dynamical behaviour of the system within the sphere and in particular at the centre of the sphere (the point) is still influenced by the static neighbourhood outside of the sphere (the set) in statistical average.} $l_0$ has been studied in a number of different glass forming models \cite{bouchaud2004adam, cavagna2007mosaic, biroli2008thermodynamic, sausset2010growing, berthier2012static, cavagna2012dynamic, cammarota2013confinement, berthier2019can}. \rp{Here we propose an analogous definition of $l_0$ that can be formulated for continuum models a follows: $l_0$ is the largest radius of a spherical region, within which the density profile in 50\% of our simulations is uniquely determined by the fixed density profile in the environment surrounding the sphere. Using this definition, we show that determining the dependence of $l_0$ on the model parameters is possible in the framework of continuum models, such as the PFC model.}

\rp{Note that the point-to-set correlation length is determined by correlations between multiple points (specifically those on the surface of a sphere with a point at the centre of the sphere) and is thus very different from the more typically considered correlation length that is obtained from two-point correlation functions such as the radial distribution function or its Fourier transform, the static structure factor \cite{HansenMcDonald}. As discussed e.g.\ in Ref.~\onlinecite{berthier2011theoretical}, two-point correlation functions are not able to distinguish clearly a glass from a liquid, but the many-body point-to-set $l_0$ does provide useful quantification.}

One may argue that the most fundamental physical theory for the structure, phase behaviour and dynamics of soft condensed matter systems is classical density functional theory (DFT) \cite{HansenMcDonald, Evans79, lutsko2010} and its dynamic extensions, dynamical density functional theory (DDFT)  \cite{HansenMcDonald, te2020classical, marconi1999dynamic, Archer2004a, archer2006dynamical, archer2009dynamical} and power functional theory \cite{schmidt2013power, schmidt2018power}. The PFC model may be viewed as a greatly simplified but very efficient version of the theory \cite{PhysRevLett.88.245701, PhysRevE.70.051605, PhysRevB.75.064107, PFCreview2012, archer2019deriving}. It has already been successfully used to study the microscopic structural aspects of homogeneous and heterogeneous crystal nucleation and grain growth \cite{doi:10.1080/14786435.2010.487476, PhysRevLett.106.195502, PhysRevLett.108.025502, C3CS60225G, PODMANICZKY201724}, amorphous precursor mediated two-step crystal nucleation \cite{PhysRevLett.107.175702}, and glass formation and ageing \cite{PhysRevLett.106.175702, PhysRevE.89.062303}. In essence, the PFC model is a statistical physics based, computationally efficient classical continuum theory capable of capturing particle-scale processes in space, whilst accessing the time scales of pattern formation during solidification (that are often orders of magnitude larger than the time scales accessible by molecular simulations). This unique feature makes the PFC an ideal tool to study the microscopic structural aspects of solid nucleation and grain growth. We build on our previous work on the amorphous nucleation precursor \cite{PhysRevLett.107.175702}, here using the PFC model to study the structural aspects of the formation of solid structures from the undercooled melt in a system that exhibits freezing into the body-centered cubic (bcc) crystalline phase and/or an amorphous structure in rapid thermal quenches.

The structure of the rest of the paper is as follows: After introducing the PFC model in Sec.~\ref{sec:2}, we briefly present the numerical technique we apply to solve the dynamical equation. The results of the numerical simulations addressing (i) the distribution of the free energy density of the bulk amorphous energy minima, (ii) the point-to-set correlation length, and (iii) the formation and stability of the solid configurations are presented and discussed in Sec.~\ref{sec:3}. The conclusions of the paper are summarised in Sec.~\ref{sec:4}.

\section{Methodology}
\label{sec:2}

\subsection{The PFC model}

The free energy functional on which the PFC model is based may be obtained by starting from the Ramakrishnan-Youssouff (RY) DFT for freezing \cite{RY1979} and making a number of further simplifying approximations. The RY-DFT is itself obtained by Taylor expanding the Helmholtz free energy functional $F[\rho]$ of the inhomogeneous system with number density $\rho(\mathbf{r})$, relative to a homogeneous reference state with density $\rho_0$, as follows:
\begin{equation}
\label{eq:2.1}
\frac{\Delta F}{k_BT}= \int  dV\,\left\{ \rho \ln \frac{\rho}{\rho_0} - \Delta \rho - \frac{\Delta \rho}{2} (c^{(2)} * \Delta \rho) \right\} \enskip ,          
\end{equation}
where $\Delta F=F[\rho]-F[\rho_0]$, $\Delta \rho(\mathbf{r})=\rho(\mathbf{r})-\rho_0$, $k_B$ is Boltzmann's constant, and $T$ is the temperature. Furthermore, $c^{(2)}(r)$ is the pair direct correlation function \cite{HansenMcDonald}, and $*$ denotes a convolution. Detailed derivations of the PFC free energy from the RY-DFT model are given e.g.\ in Refs.~\onlinecite{PFCreview2012, archer2019deriving}. It starts with the introduction of the relative density difference $n({\mathbf{r}}) \equiv \frac{\rho-\rho_0}{\rho_0}$. Using this in the local terms of Eq.~(\ref{eq:2.1}) yields:
\begin{equation}
\label{App1}
\rho \ln \frac{\rho}{\rho_0}-\Delta\rho = \rho_0 \left[(1+n) \ln(1+n) - n \right] \enskip .
\end{equation}
Taylor expanding Eq.~(\ref{App1}) up to the $4^{th}$ order results in:
\begin{equation}
\label{App2}
(1+n)\ln(1+n)-n\approx \frac {n^2}{2} - \frac{n^3}{6} + \frac{n^4}{12} \enskip .
\end{equation} 
Furthermore, the convolution part of Eq.~(\ref{eq:2.1}) contains the correlation function $c^{(2)}(r)$. The Fourier transform of $c^{(2)}(r)$ (denoted by $\tilde{c}^{(2)}(k)$ henceforth) has multiple peaks, which results in finite length-scale selection. To simplify $\tilde{c}^{(2)}(k)$ together with keeping length-scale selection, we replace $\tilde{c}^{(2)}(k)$ by its Taylor expansion up to the $4^{th}$ order:
 \begin{equation}
\label{App3}
\tilde{c}^{(2)}(k) \approx C_0+C_2 k^2 + C_4 k^4 \enskip .
\end{equation}
Using Eqs.~(\ref{App2}) and (\ref{App3}) in Eq.~(\ref{App1}), and adding the $0^{th}$ order contributions from the $3$ and $4$-point correlations (corrections from beyond the RY-DFT functional) yields the following dimensional PFC free energy functional:
\begin{equation}
\label{App4}
\frac{\Delta F_{PFC}}{\rho_0 {k_{B}T}}= \int dV \left\{ \frac{n^2}{2} - a\frac {n^3}{3} + b \frac {n^4}{4} - \frac{\rho_0} {2} n\,\hat{\mathcal{L}}\,n\right\} \enskip ,        
\end{equation}
where $\hat{\mathcal{L}}=C_0-C_2 \nabla^2 + C_4 \nabla^4$. There are $8$ parameters in Eq.~(\ref{App4}) ($\rho_0, T, a, b, C_0, C_2, C_4$ and the average density $\bar{n}\equiv \frac{1}{V}\int n dV$). However, these can be greatly reduced in number by the following non-dimensionalisation process: First we introduce the function $Q(k) \equiv [\mathcal{L}(k_0)-\mathcal{L}(k)]/[\mathcal{L}(k_0)-\mathcal{L}(0)]$, where ${\mathcal{L} (k)} \equiv  C_0+C_2 k^2 + C_4 k^4$, and $k_0=\sqrt{C_2/(2|C_4|)}$ is the position of the peak of $\mathcal{L}(k)$. Consequently, $\mathcal{L} (k) = L - v \, {{Q} (k)}$, where  $L=\mathcal{L} (k_0)$ and $v= L- C_0$. Using this in (\ref{App4}) results in:  
 \begin{equation}
\label{App7}
\frac{\Delta F_{PFC}}{\rho_0 k_BT}= \int dV \left\{n\frac{1-\rho_0 {(L - v\,\hat{Q})}}{2} n-a\frac{n^3}{3} + b\frac{n^4}{4}\right\} \enskip ,         
\end{equation}
where $\hat{Q} = c_0-c_2\nabla^2+c_4 \nabla^4$ with constants $c_0$, $c_2$ and $c_4$ depending on $C_0$, $C_2$ and $C_4$. $F$, $n$ and $\mathbf{r}$ are non-dimensionalised by introducing the units $A$, $X$ and $\lambda$, respectively. Choosing the length scale $\lambda \equiv 1/k_0$ results in
\begin{equation}
\label{App8}
\hat{Q} = (1+\nabla'^2)^2
\end{equation}
for any $C_0$, $C_2$ and $C_4$. Finally, choosing $X \equiv \sqrt{b \rho_0/v}$ and $A \equiv \frac {Tv^2 \rho^3_0 k_B}{b k^3_0}$ yields:
\begin{equation}
\label{App9}
 \mathcal{F}= \int dV^{\prime} \left\{\Phi\frac{(1+ {\nabla^{\prime}}^2)^2 - \alpha}{2} \Phi - \tau \frac{\Phi^3}{3} + \frac{\Phi^4}{4}\right\} \enskip ,          
\end{equation}
where $\Phi = n/X$ and $\mathcal{F} = \Delta F_{PFC}/A$ are the dimensionless density and free energy, respectively. Furthermore, the model parameters read:
\begin{eqnarray}
\alpha &=& =1+ \left(\frac {4 C_4}{C^2_2}\right) \frac {1-\rho_0 C_0}{\rho_0} \\
\tau &=& \frac {2a}{C_2} \sqrt {\frac {|C_4|}{b \rho_0}} \enskip .
\end{eqnarray} 
The cubic term in Eq.~(\ref{App9}) can be eliminated by introducing the new variable $\psi \equiv \Phi-\tau/3$. Substituting this into Eq.~(\ref{App9}), and \rp{omitting the linear term $A\psi+B$ in the integrand (which has no effect on the phase diagram or the dynamics of the system)} yields the ``minimal form'' of the PFC energy functional:
\begin{equation}
\label{eq:2.2}
\mathcal{F}= \int dV \, \left\{ \psi \frac{(1+\nabla^2)^2-\gamma}{2}\psi + \frac {\psi^4}{4} \right\} \enskip ,       
\end{equation}
where $\gamma = \alpha+\tau^2/3$, which is the only model parameter in addition to the average density $\bar{\psi} = \bar{\phi}-\tau/3$. \rp{Comparing Eqs.~(\ref{App9}) and (\ref{eq:2.2}) also shows that the sole effect of including a cubic term in the free energy functional is a shift in the phase diagram.} 

The time evolution of over-damped Brownian systems (such as colloidal suspensions) is described by DDFT \cite{HansenMcDonald, te2020classical, marconi1999dynamic, Archer2004a} and together with the PFC model simplifications, this free energy minimising dynamics becomes
\begin{equation} 
\label{eq:2.3}
\frac{\partial \psi}{\partial t} = \nabla \cdot \left\{ M(\mathbf{r})\nabla \left(\frac{\delta \mathcal{F}}{\delta \psi} \right)\right\},
\end{equation}
where $\frac{\delta \mathcal{F}}{\delta \psi}=\left[(1+{\nabla}^2)^2 - \gamma\right] \psi +\psi^3$ is the first functional derivative of $\mathcal{F}$ with respect to $\psi$, and $M(\mathbf{r})\geq0$ is a generic non-constant mobility function. \rp{It can be shown that Eq.~(\ref{eq:2.3}) defines a dynamics that leads to the free energy decreasing monotonically in time, i.e.
\begin{equation}
\frac{dF[\psi(\mathbf{r},t)]}{dt} \leq 0,
\end{equation}
for any $M(\mathbf{r}) \geq 0$ and periodic boundary conditions.} We introduce the spatial dependence into the mobility $M(\mathbf{r})$ in order to calculate the point-to-set correlation length $l_0$ of the amorphous phase. In these calculations we set $M(\mathbf{r})=0$ outside the spherical region where $|\mathbf{r}|\gtrsim r_0$ and $M(\mathbf{r})=M_0$, a constant, within the spherical region where $|\mathbf{r}|\lesssim r_0$ -- for details see Sec.~\ref{sec:3.2}, below. The largest value of $r_0$ for which the density profile within the sphere remains determined by the profile of the surrounding amorphous material in the majority of our samples defines the point-to-set correlation length $l_0$. In all other calculations, we set $M(\mathbf{r})=M_0$ for all positions $\mathbf{r}$ in the system.

For numerical reasons, we introduce the variable $\phi(\mathbf{r}, t)\equiv\psi(\mathbf{r}, t)-{\overline \psi}$, where $\overline \psi$ is the average density of the system, which is preserved by Eq.~(\ref{eq:2.3}). Using this variable transformation in Eq.~(\ref{eq:2.3}) gives
\begin{equation}
\label{eq:2.4}
\frac{\partial \phi}{\partial t} = \nabla \cdot \left[ M(\mathbf{r}) \nabla \left\{\left[(1+{\nabla}^2)^2 - \epsilon\right] \phi +(3 \overline \psi)\phi^2 +\phi^3 \right\}\right],             
\end{equation}
where \rp{$\epsilon =  \gamma - 3 \overline\psi^2$}, and the spatial average of $\phi(\mathbf{r}, t)$ is zero. Eq.~(\ref{eq:2.4}) is solved numerically to find local minima of the free energy at fixed model parameters $\overline\psi$ and $\gamma$.

\rp{A final point worth mentioning regarding Eq.~(\ref{eq:2.3}) is that in the derivation of DDFT, this equation is obtained by averaging over all realisations of the thermal noise for a system of interacting Brownian particles \cite{HansenMcDonald, te2020classical, marconi1999dynamic, Archer2004a, Archer2004}. Thus, in principle, all the effects of thermal fluctuations are incorporated into the theory already. However, in practice, we must always use an approximate (non-equilibrium) free energy functional $\cal{F}$ in conjunction with this equation, so we inevitably neglect some fluctuation effects. In our work here, as discussed below in Sec.~\ref{sec:3.1}, we do include randomness into our initial conditions, but we do not add a stochastic term to Eq.~(\ref{eq:2.3}), which would incorporate further fluctuation effects over the time evolution, since this would in effect be double-counting some of the effects of thermal noise.}

\subsection{Numerical method}

To solve Eq.~(\ref{eq:2.4}) numerically, we apply an operator-splitting based quasi-spectral semi-implicit time-stepping scheme \cite{TEGZE20091612}, as follows: Consider the partial differential equation 
\begin{equation}
\label{eq:2.5}
\frac{\partial \phi (\mathbf r, t)}{\partial t} = \hat O[\phi (\mathbf r, t)],             
\end{equation}
where the right-hand side $\hat O[.]$ is a function of time, space, $\phi, \nabla\phi, \nabla^2\phi$, and so on. The development of the time-stepping scheme starts by adding and subtracting a linear term $ \hat s [\phi (\mathbf{r}, t)]$ (called the splitting) as
 \begin{equation}
\label{eq:2.6}
\frac{\partial \phi (\mathbf r, t)}{\partial t} = \hat O[\phi (\mathbf r, t)]+ \hat s[\phi (\mathbf r, t)] - \hat s[\phi (\mathbf r, t)] \enskip ,              
\end{equation}
which is still identical to Eq.~(\ref {eq:2.5}). The time derivative is then discretised using the forward Euler scheme, while the first two terms on the right-hand side are taken at time $t$. Finally, the last term on the right-hand side is taken at time $t+\Delta t$, where $\Delta t$ is the time-step, yielding
\begin{equation}
\label{eq:2.7}
\frac {\phi^{t+\Delta t} (\mathbf{r}) - \phi^t (\mathbf{r})}{\Delta t} = \hat O[\phi^t (\mathbf{r})]+ \hat s[\phi^t(\mathbf{r})] - \hat s[\phi^ {t+\Delta t} (\mathbf{r})] \enskip .
\end{equation}
Imposing periodic boundary conditions, Fourier transforming and then re-arranging Eq.~(\ref{eq:2.7}) yields the following time-stepping scheme:
\begin{equation}
\label{eq:2.9}
 \tilde \phi_{\mathbf k}^{t+\Delta t} = \tilde \phi^{t}_{\mathbf k} +\frac {\Delta t}{1+\Delta t \, s(k)} \, \mathcal{F}_{\mathbf k} \left\{ \hat O[\phi^t (\mathbf{r})]\right\} \enskip ,
\end{equation}
where $\tilde \phi^{t}_{\mathbf k}$ is the Fourier coefficient of $\phi^t(\mathbf r)$, while $s(k)$ is related to the Fourier transform of $\hat{s}[\delta(\mathbf{r})]$, where $\delta(\mathbf{r})$ is the Dirac delta. The main benefit of Eq.~(\ref {eq:2.9}) is that it is almost unconditionally stable for a suitably chosen splitting function $s(k)$.

\begin{figure}
    \includegraphics[width=1.0\linewidth]{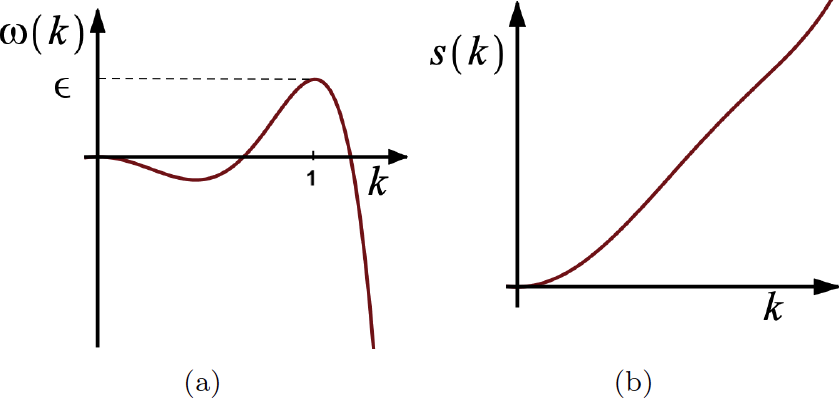}
  \caption{Dispersion relation (a) and splitting operator (b) for the PFC model.}
 \label{fig:1}
 \end{figure}

To find the function $s(k)$ for which Eq.~(\ref{eq:2.9}) is stable for reasonably large $\Delta t$ values we note that the right-hand side of Eq.~(\ref {eq:2.5}) [see Eq.~(\ref {eq:2.4})] reads:
\begin{equation}
\label{eq:2.10}
\hat O[\phi(\mathbf{r})]=\nabla \cdot \left\{ M \nabla \left[\hat{\mathcal{L}} (\phi) + f_{nl} (\phi)  \right]\right\},
\end{equation}
where $\hat{\mathcal{L}}=(1+\nabla^2)^2 -  \epsilon$ is a linear operator, and  $f_{nl} (\phi) = (3 \overline \psi)\phi^2 +\phi^3$ a non-linear function. To find the splitting operator $\hat s\left[\phi\right]$ we need to expand Eq.~(\ref{eq:2.10}). Since $M$ is non-constant in general, we add and subtract a constant $M_0$ to $M$ (where $M_0 > 0$ is an upper bound to $M$), thus, giving
\begin{eqnarray} 
\label{eq:2.11}
\hat O[\phi]&=& \nabla\cdot\left\{M\nabla\hat{\mathcal{L}} \left[\phi\right] \right\}+ \nabla \cdot \left\{M\nabla f_{nl} (\phi)\right\} \nonumber\\
&=&\nabla \cdot \left\{\left(\Delta M+M_0\right) \nabla \hat{\mathcal{L}} \left[\phi\right] \right\} + \nabla \cdot \left\{M\nabla f_{nl} (\phi)\right\} \nonumber\\
&=& M_0\nabla^2 \hat{\mathcal{L}} \left[\phi\right]+\nabla \cdot \underline{\mathcal J}, 
\end{eqnarray}
where \rp{$\underline{\mathcal J}= \Delta M\nabla\hat{\mathcal{L}}[\phi]+M \nabla f_{nl} (\phi)$ and $\Delta M=M-M_0$}. The Fourier transform of the linear part $M_0 \, \nabla^2 \, \hat{\mathcal{L}} (\phi)$ is $\omega(k)\phi_k$, where $\omega(k)$ is the dispersion relation, which reads:
\begin{equation}
\label{eq:2.12}
 \omega (k)= M_0\left(-k^2\right) \left[(1-k^2)^2 -  \epsilon\right].
\end{equation}
 Since $\omega (k)$ gives the relaxation rate of perturbations as a function of the wave number $k$, only modes around $k\approx1$ grow for $\epsilon > 0$ -- see Fig.~\ref{fig:1}(a).
To stabilise the growth of these modes for $0<\epsilon<1$ we choose the splitting operator as 
\begin{eqnarray} 
\label{eq:2.13}
 s(k) &\equiv& -\left[\omega (k) + M_0k^2\right]\nonumber\\
 &=& M_0 k^2 \left[(1-k^2)^2 +1-  \epsilon \right] > 0,
\end{eqnarray}
for $k>0$; see Fig.~\ref{fig:1}(b).

\section{Results}
\label{sec:3}

\subsection{Free energy density}
\label{sec:3.1}

We first investigated the statistical properties of the bulk amorphous phase by solving Eq.~(\ref{eq:2.4}) with $M(\mathbf{r})=1$ (i.e.\ a spatially uniform mobility and without loss of generality setting $M_0 = 1$) over the time interval [$0, T$] and on various different sized 3-dimensional spatial domains $[0, L]^3$ with periodic boundary conditions. The grid spacing of the uniform computation grid and time step are fixed as $\Delta x=2/3$ and $\Delta t=1/4$, respectively. The initial condition is chosen as
\begin{equation}\label{eq:chi_U}
\rp{\phi(\mathbf{r}, 0)=\chi U(\mathbf{r}),}
\end{equation}
where $U(\mathbf{r})$ is a random \rp{field with uniform distribution on $[-1, 1]$. The parameter $\chi$ sets the amplitude of the random perturbations. The initial state therefore corresponds to that of a uniform liquid, with the superimposed noise modelling thermal density fluctuations and the amplitude of these is determined by the parameter $\chi$. By choosing $\gamma$ and $\bar{\psi}$ corresponding to a point in the phase diagram where the crystal is the equilibrium phase, but choosing the initial state to correspond to that of a liquid, we are modelling a rapid (in fact, instantaneous) quench of the system from a high temperature molten state down to a lower temperature.} Thus, roughly speaking, larger values of $\chi$ correspond to deeper temperature quenches \rp{(i.e.\ moving from higher temperature states to the selected final state)}, while smaller values of $\chi$ correspond to shallower temperature quenches \rp{(descending from lower temperature states)}.

\begin{figure}
 \includegraphics[width=1.0\linewidth]{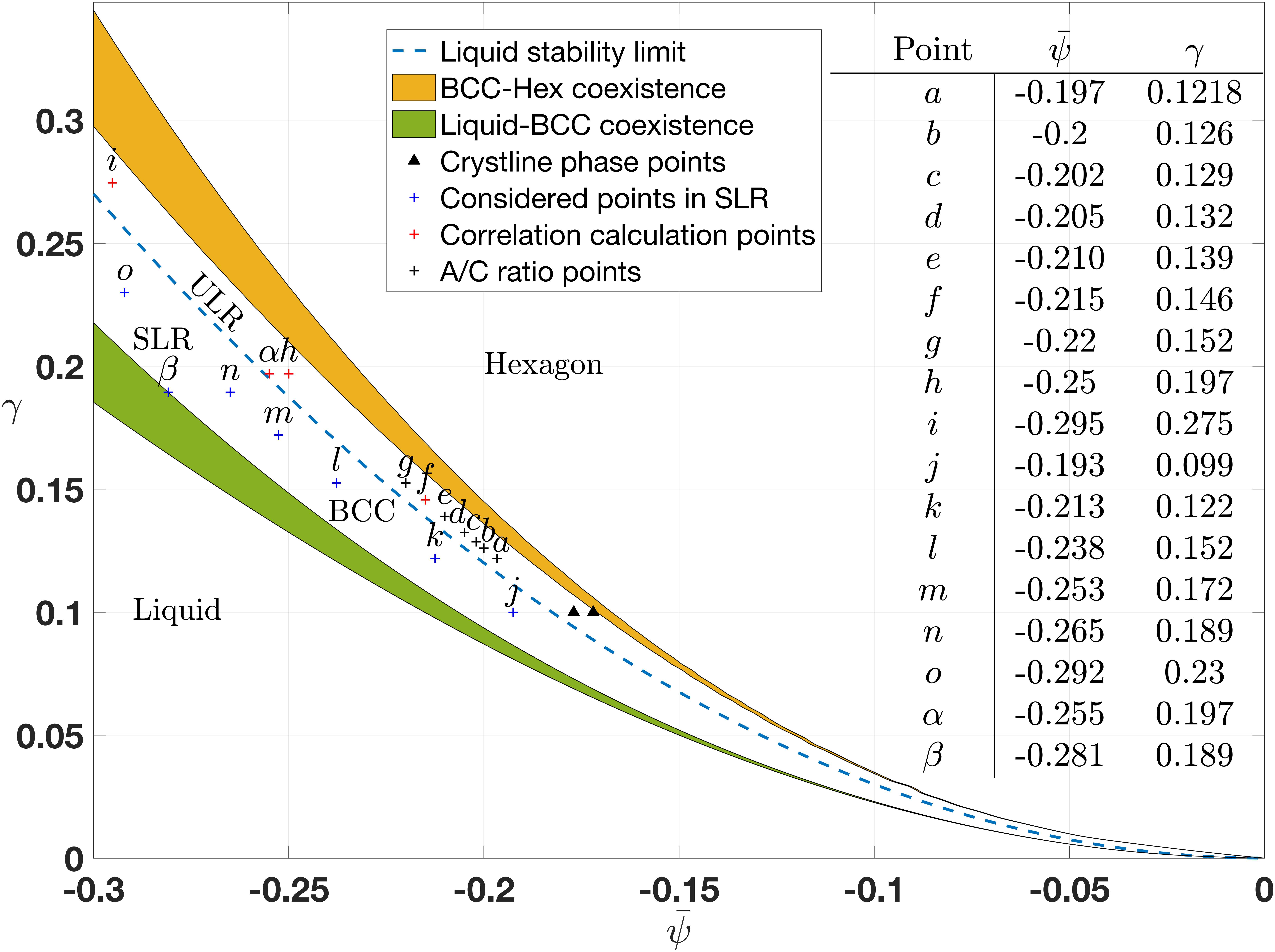}
  \caption{Phase diagram of the PFC model in the average density $\bar{\psi}$ versus inverse-temperature-like parameter $\gamma$ plane. The uniform liquid is the equilibrium state in the bottom left hand region of the phase diagram. On increasing either $\gamma$ or $\bar{\psi}$, a transition to the BCC crystal occurs. The green region denotes the region of bulk phase coexistence between the liquid and the BCC. The model also exhibits other periodic phases such as \rp{columnar hexagons and lamellae}, that are not of interest here. \rp{Above the linear stability limit (dashed line) the liquid is linearly unstable; we refer to this as the unstable liquid region (ULR), whereas below, in the SLR, the liquid is linearly stable.} The symbols and the corresponding letters denote the locations of state points at which the numerical simulations are performed -- see below.}
 \label{fig:3.0}
 \end{figure}
 \begin{figure}
  \centering
 \includegraphics[width=1.0\linewidth]{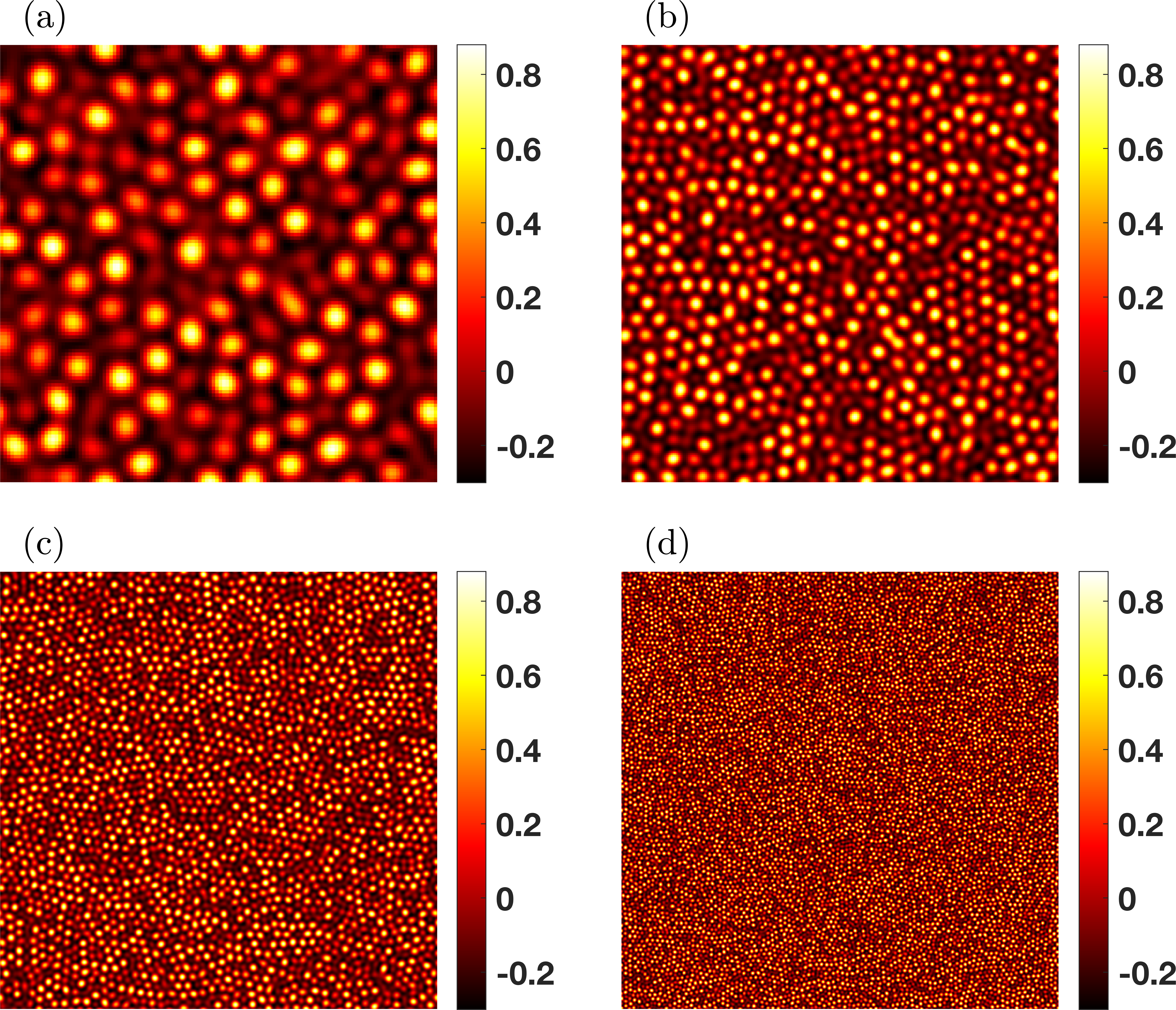}
  \caption{Two-dimensional cross sections of converged 3-dimensional amorphous configurations for system sizes (a) $N=128$, (b) $N=256$, (c) $N=512$ and (d) $N=1024$, at state point $h$ in Fig.~\ref{fig:3.0}.}
 \label{fig:3}
  \end{figure}
 \begin{figure}
  \centering
 \includegraphics[width=1.0\linewidth]{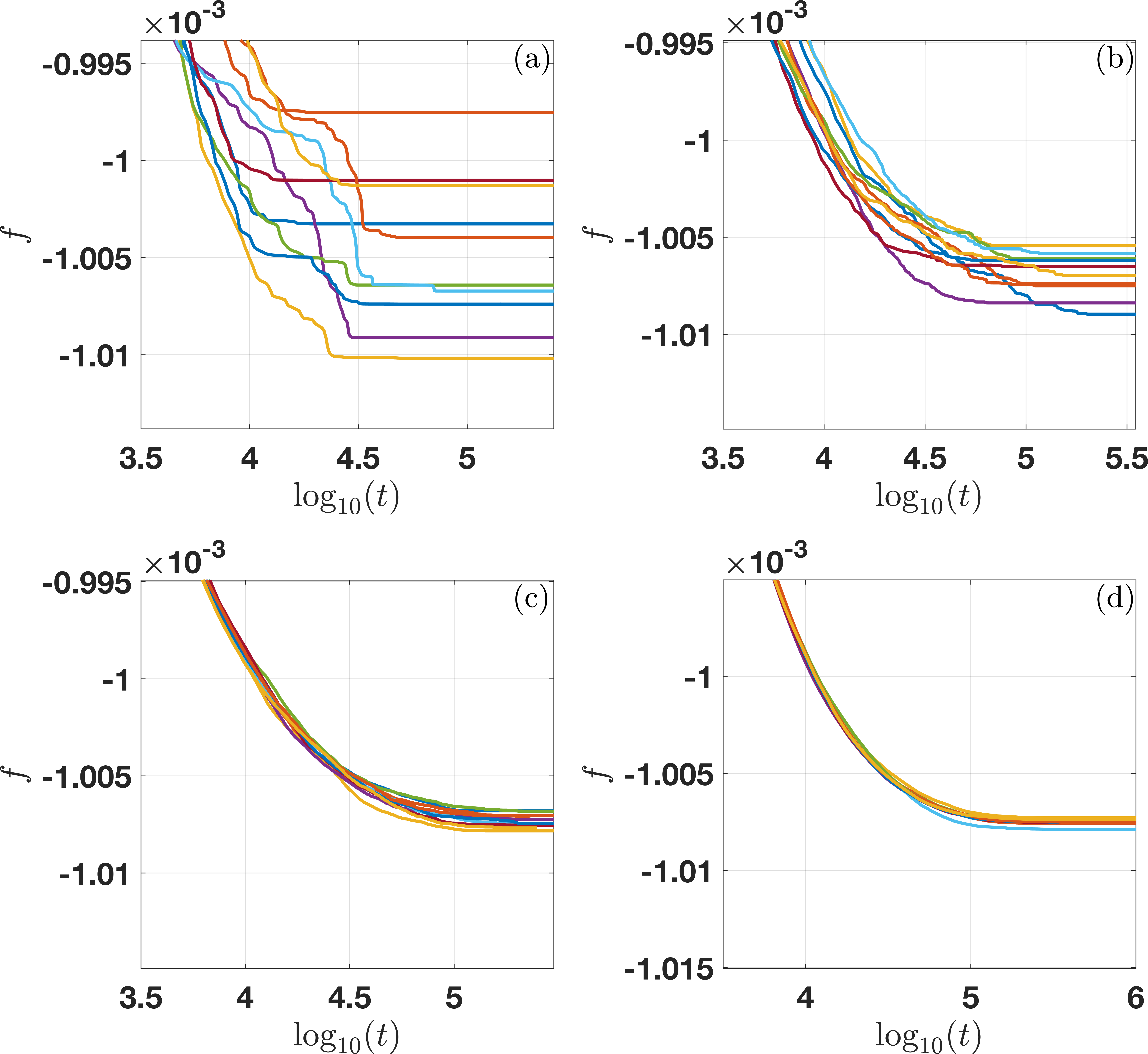}
  \caption{Time evolution of the free energy density for 10 different initial conditions at $\chi = 10 \overline{\psi}$ for system sizes (a) $N=128$, (b) $N=256$, (c) $N=512$ and (d) $N=1024$.}
 \label{fig:3.1}
 \end{figure}

A portion of the bulk equilibrium phase diagram is displayed in Fig.~\ref{fig:3.0} -- for the full phase diagram see e.g.\ Ref.~\onlinecite{PFCreview2012} and references therein. Initially, we consider the case where the model parameters are $\gamma=0.197$ and $\bar\psi=-0.25$ (indicated by point $h$ on the phase diagram in Fig.~\ref{fig:3.0}), at which the homogeneous (liquid) phase $\phi(\mathbf{r})=0$ is an unstable stationary point of the free energy and the global minimum of the free energy is the BCC crystal phase. Since at this state point the liquid is linearly unstable, a barrierless phase transformation takes place, i.e., the liquid phase evolves to a non-uniform density minimum of $F$ for any $\chi>0$. In our simulations, we considered 4 different system sizes $L=N\Delta x$, with $N=128, 256, 512$, and $1024$, to study the effect of the system size (with periodic boundary conditions) on the converged numerical solution. For each system size, we repeat the simulation multiple times starting from different initial conditions (i.e.\ different realisations of the noise field), initially with large amplitude $\chi=10 \bar\psi$. At each system size, the perturbed liquid evolves into either the bulk amorphous or bulk crystalline states. In Fig.~\ref{fig:3} we display 2-dimensional cross sections of typical 3-dimensional bulk amorphous configurations.

\begin{figure}
\centering
\includegraphics[width=0.99\linewidth]{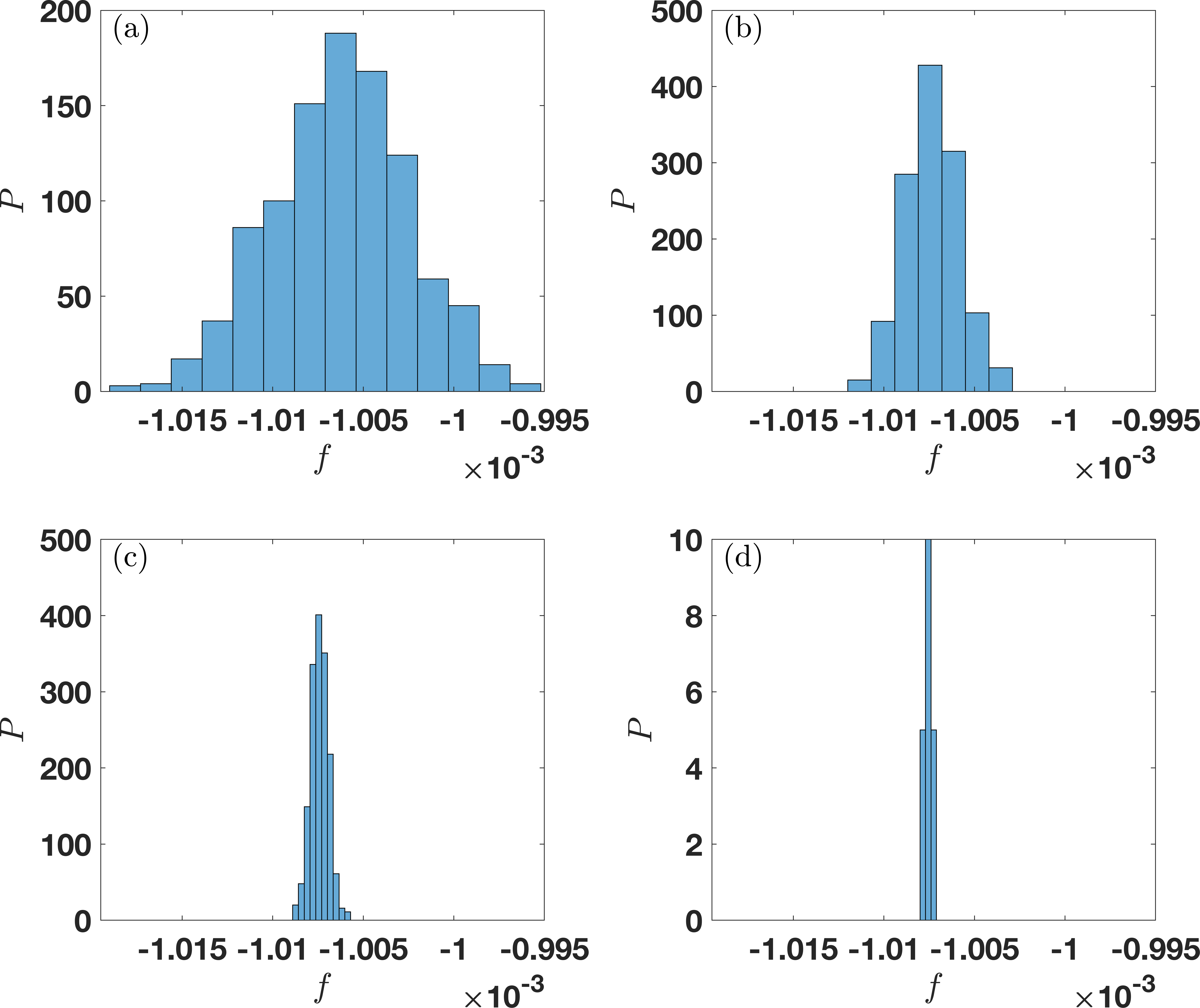}
\caption{Histograms for the different values of the free energy density $f$ as a function of system size $N$ (the system volume is $L^3$, with $L=N\Delta x$): (a) $N=128$ (1000 simulations), (b) $N=256$ (1269 simulations), (c) $N=512$ (1611 simulations)  and (d) $N=1024$ (20 simulations).}
\label{fig:3.2}
\end{figure}

\begin{figure}
\centering
\includegraphics[width=1.0\linewidth]{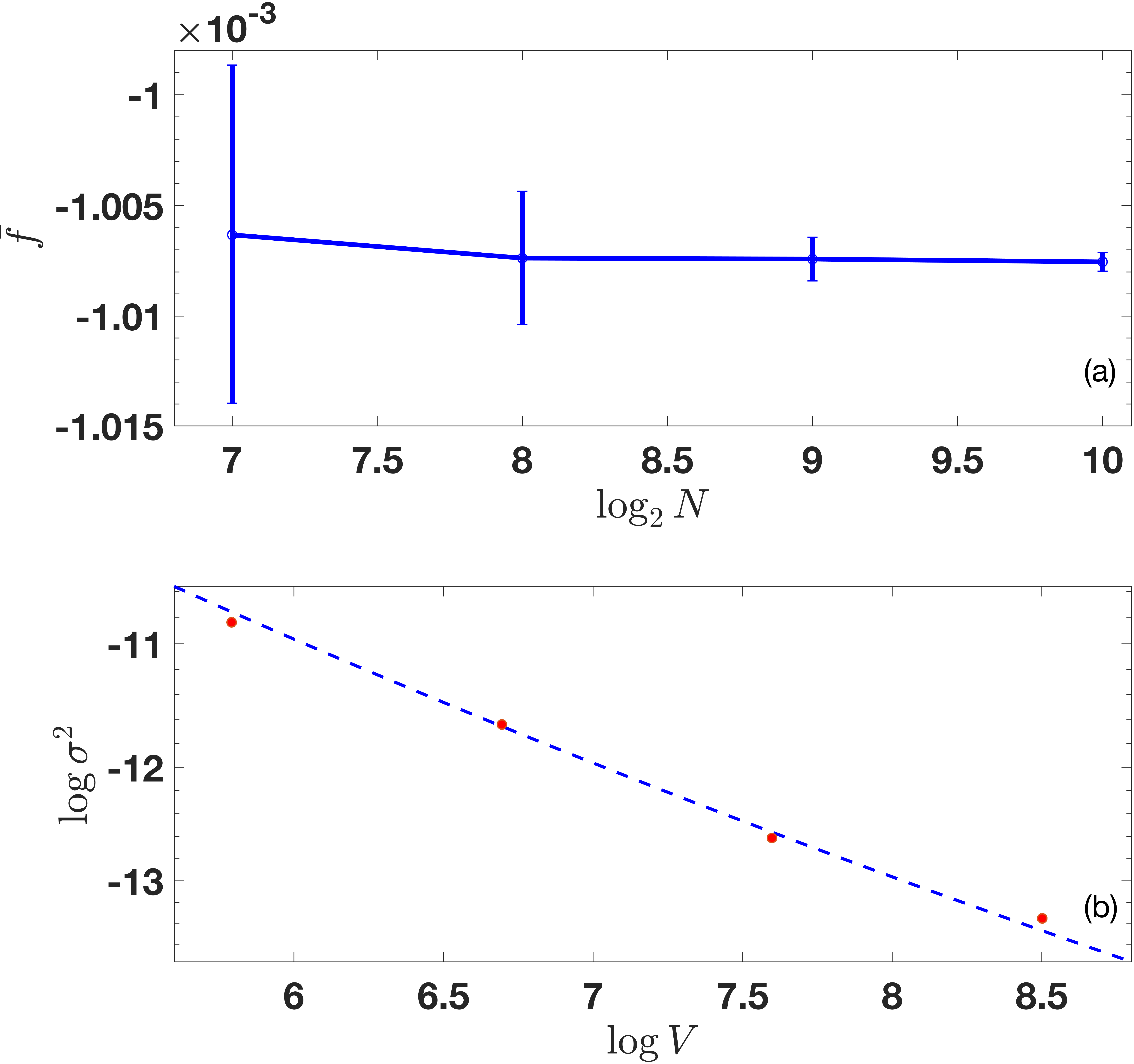}
\caption{Mean free energy density $\bar{f}$ and the logarithm of its variance, $\log\sigma^2$, as a function of system size. In (a) $\bar{f}$ is plotted as a function of $\log_2N$ and the error bars indicate $\bar{f}\pm2\sigma$. In (b) the symbols show $\log\sigma^2$ as a function of the system volume $V=(N\Delta x)^3$. The dashed straight line has gradient $=-1$. }
\label{fig:3.3}
\end{figure}

In all simulations, the free energy \rp{decreases over time and we run our simulations for a sufficiently long time that the value of the free energy for each system converges. This can be seen from Fig.~\ref{fig:3.1}, where we plot the time variation of the free energy density relative to that of the liquid at the same average density,
\begin{equation}
f \equiv \frac{\mathcal{F}[\bar{\psi}+\phi]-\mathcal{F}[\bar{\psi}]}{V},
\end{equation}
where $\mathcal{F}[\psi]$ is defined in Eq.~(\ref{eq:2.2}) and $V=L^3=(N\Delta x)^3$ is the volume. Fig.~\ref{fig:3.1} shows how $f$ converges over time. For amorphous final} states, numerous possible values of $f$ are obtained, corresponding to the multiple different (local minima) disordered structures that the system can evolve into. We find a distribution $P(f)$ of observed free energy density values that has a variance $\sigma^2$ that decreases with increasing system size -- see Figs.~\ref{fig:3.2} and \ref{fig:3.3}. Thus, in the thermodynamic limit (\rp{$V\to\infty$, for fixed average density $\bar{\psi}$}) we can expect to find a well-defined value of $f$ for the amorphous state. The decrease in the variance $\sigma^2$ as $N$ increases is due to (i) the decreasing effect of the periodic boundary conditions (larger systems are less confined by the boundary conditions), and (ii) a finite correlation length of the forming pattern (i.e.\ a finite value of the point-to-set correlation length), that we return to in detail below. 

\begin{figure}
\centering
\includegraphics[width=1.0\linewidth]{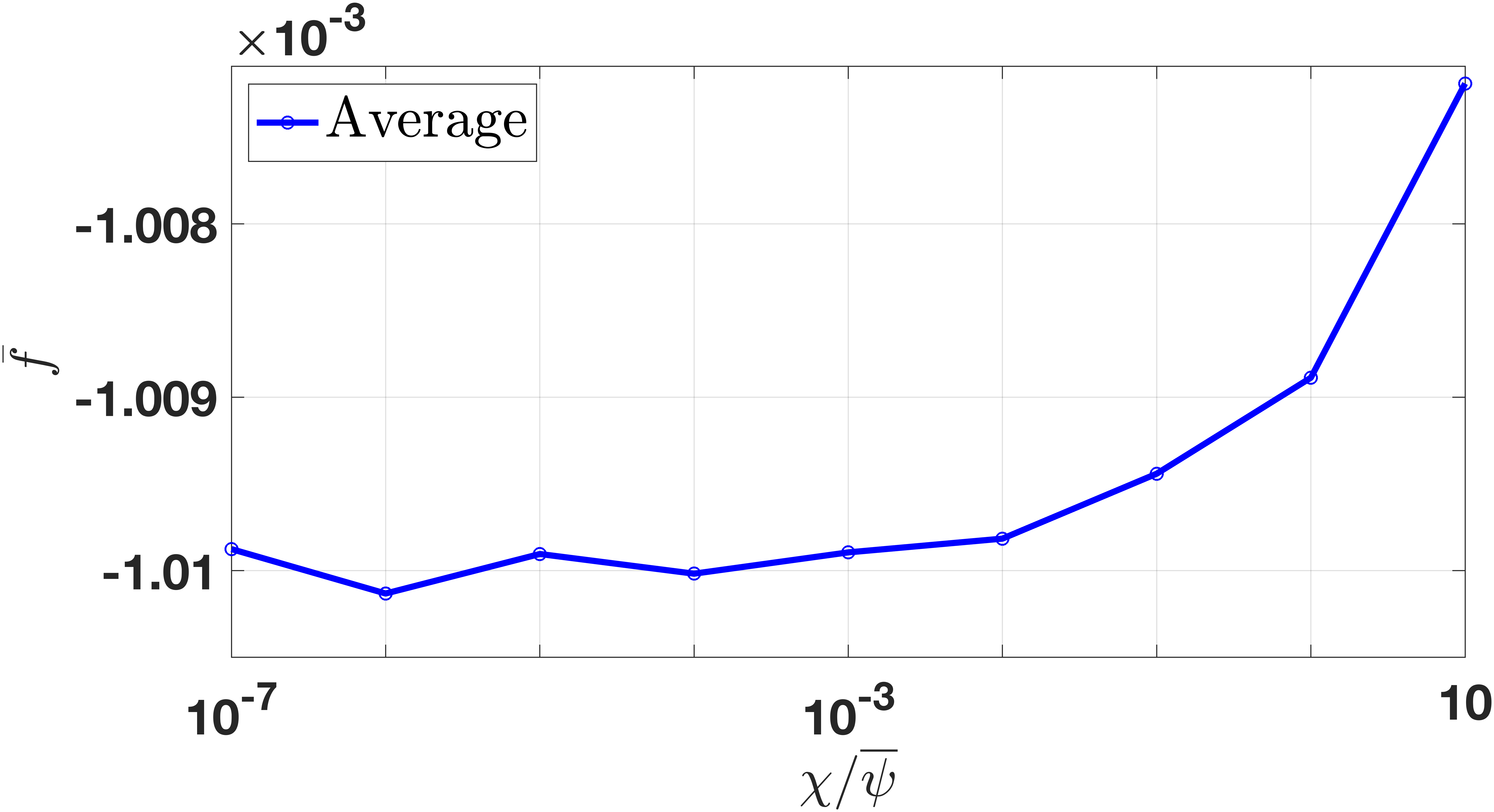}
\caption{Mean free energy density $\bar{f}$ as a function of the amplitude of the noise in the initial conditions $\chi$, at point $h$ in the phase diagram and for fixed $N=512$.}
\label{fig:3.32}
\end{figure}

In Fig.~\ref{fig:3.3} we plot the mean free energy density $\bar{f}$ and the variance $\sigma^2$ as the system size $N$ is varied. Figure \ref{fig:3.3}(a) shows that once the system size $N$ is large enough that finite size effects become negligible, then the mean value $\bar{f}$ is independent of $N$.

In Fig.~\ref{fig:3.3}(b) we plot the logarithm of the variance, $\log\sigma^2$, as a function of \rp{the logarithm of the volume, $\log V$}. If we consider each configuration of the system to be made up of finite-sized building blocks of volume $V_0\approx \tfrac{4}{3}\pi l_0^3$ (recall, $l_0$ is the point-to-set correlation length) with the energy density $f$ in each block being independent of the value in the surrounding blocks and drawn from a distribution with standard deviation $\sigma_0$, then the central limit theorem indicates that as the system size is increased, the variance of the energy density $\sigma^2\to\sigma_0^2/n$, where $n\approx V/V_0$ is the number of blocks in the system. In other words, as the system volume $V$ is increased, this hypothesis indicates that the variance should decrease with system size as $\sigma^2\propto1/V$. The straight line logarithmic plot in Fig.~\ref{fig:3.3}(b) with gradient $\approx-1$ confirms this picture of the amorphous structures that are formed. The existence of a finite correlation length $l_0$ is also consistent with the our observation that as the system size is increased, $\bar{f}$ tends to a unique value. This might be taken to indicated that in the thermodynamic limit the amorphous phase can be considered to be an well-defined \rp{metastable} phase of the system. However, this is not the whole story. When we repeat the above calculations for different values of the noise amplitude in the initial condition $\chi$, we find that $\bar{f}$ depends very slightly on $\chi$, as can be seen in Fig.~\ref{fig:3.32}. We vary $\chi$ ranging from $10^{-7}\bar{\psi}$ up to $10\bar{\psi}$. The results in Fig.~\ref{fig:3.32} show that $\bar{f}$ only varies a little as $\chi$ is increased, with the maximal relative difference between observed values being only 0.3\%. Nonetheless, these (albeit small) variations indicate that there is an evolution history dependence of the free energy of the amorphous phase, which is consistent with what is observed more generally in glassy systems \cite{berthier2011theoretical}.

\subsection{Point-to-set correlation length}
\label{sec:3.2}

 \begin{figure}
 \includegraphics[width=0.99\linewidth]{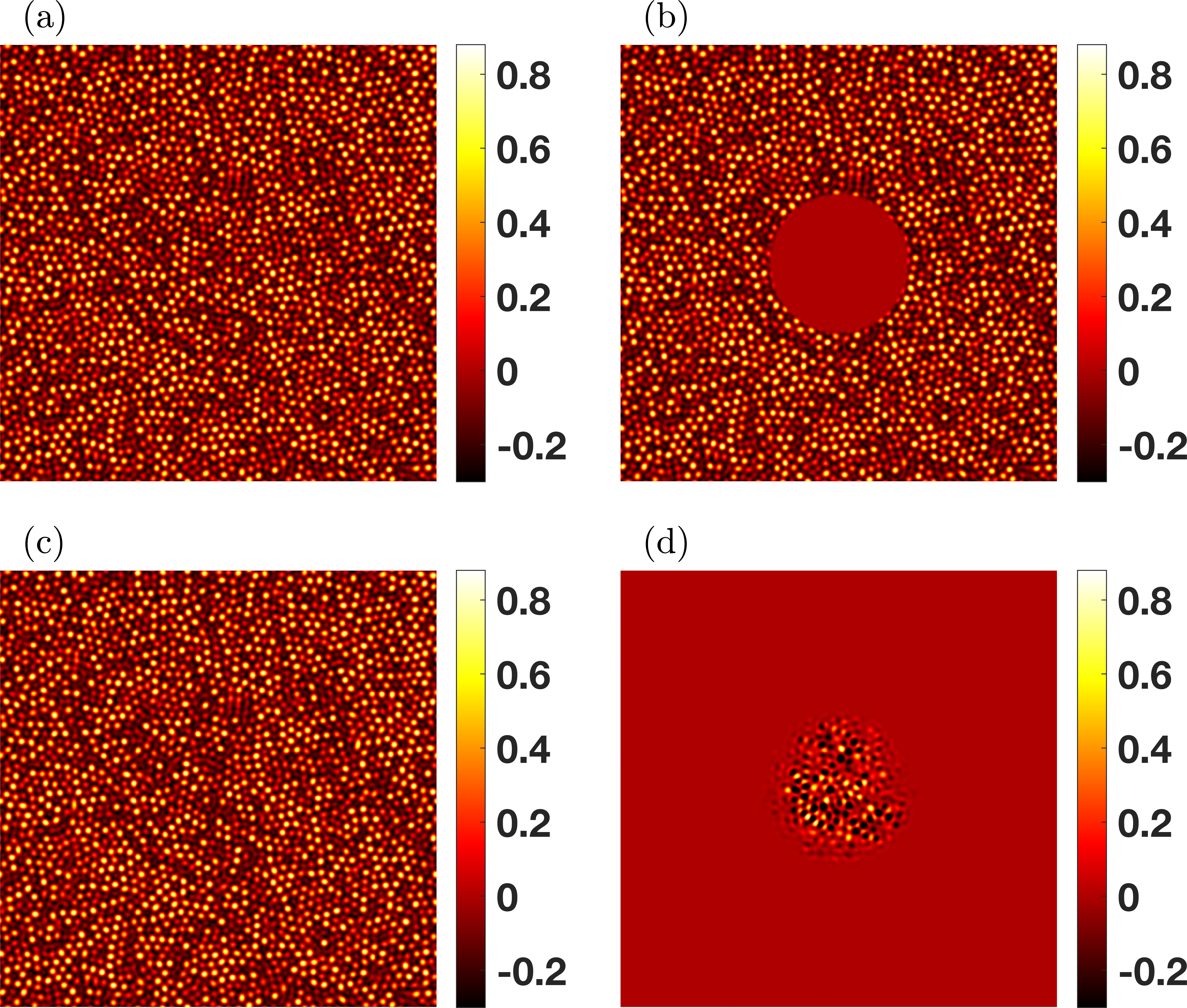}
    \caption{Illustration of how we determine the point-to-set correlation length $l_0$ in the bulk amorphous state. In (a) we display a 2-dimensional cross section of a reference bulk amorphous state of size $N=512$. We then replace the amorphous density distribution by that of the liquid within a spherical region of radius $r_0$ to create the profile displayed in (b). The profile (b) is then used as an initial condition and evolved forward in time while constraining the region outside the sphere to remain unchanged, \rp{until} the system converges, forming the state displayed in (c). In (d) we display the difference between the \rp{fields shown} in (c) and (a). This is repeated for a range of $r_0$ and for a selection of different initial states (a). $l_0$ is defined as the largest value of $r_0$ for which \rp{in half of our simulations} the final state (c) still returns to the initial state (a) (i.e.\ where the difference (d) is zero everywhere) in \rp{50\%} of our simulations. \rp{Note that in example displayed here, we have $r_0>l_0$.}}
    \label{fig:3.21}
\end{figure}

We now analyse in more detail the density correlations in the amorphous structures formed by the PFC model. In particular, we determine the point-to-set correlation length \cite{berthier2011theoretical} for converged bulk amorphous solutions of Eq.~(\ref{eq:2.4}) \rp{as follows: First we take a converged bulk amorphous configuration as a reference state -- see Fig.~\ref{fig:3.21}(a). We then replace the amorphous solid pattern by a constant value (corresponding to the liquid) within a spherical volume of radius $r_0$, as illustrated in Fig.~\ref{fig:3.21}(b). The density distribution is:}
\begin{equation}
\label{eq:3.1}
 \phi(\mathbf{r},0)\equiv\left[1-h(\mathbf{r})\right]\phi^*(\mathbf{r})+h(\mathbf{r}) \phi_0,
\end{equation}
\rp{where $\phi^*(\mathbf{r})$ is the converged bulk amorphous solution (the reference state), while}
\begin{equation}
\label{eq:3.2}
 h(\mathbf{r})= \frac {1-\tanh\left[(r-r_0)/\lambda\right]}{2}.
\end{equation}
\rp{Here, $r$ is the distance from the centre of the simulation domain, $r_0$ is the radius of the sphere where $\phi^*(\mathbf{r})$ is replaced by $\phi_0$ (constant, representing the liquid)}, and $\lambda$ the characteristic thickness of the transition zone between the \rp{liquid} and the bulk amorphous (we choose $\lambda=2\Delta x$ \rp{in our simulations}); see Fig.~\ref{fig:3.21}(b). Furthermore, setting 
\begin{equation}
\label{eq:3.3}
 \phi_0\equiv\frac{\int_\Omega dV \left\{\left[h(\mathbf{r})-1\right]\phi^*(\mathbf{r})\right\}}{\int_\Omega dV h(\mathbf{r})}
\end{equation}
in Eq.~(\ref{eq:3.1})\rp{, where $\Omega$ is the system domain,} ensures $\langle\phi(\mathbf r,t)\rangle\equiv\frac{1}{V}\int_\Omega dV\phi(\mathbf r,t)=0$, so that the average density of the system does not change, i.e.\ so that the location of the state in the phase diagram remains the same  \rp{[recall that $\phi(\mathbf{r}, t)=\psi(\mathbf{r}, t)-{\overline \psi}$ and see also Eq.~(\ref{eq:2.4})]. Using $\phi(\mathbf{r},0)$ in Eq.~(\ref{eq:3.1}) as initial condition, the} system is then allowed to evolve in time in a manner that the unaltered part of the domain is kept static. This is done by choosing \rp{
\begin{equation}
M(\mathbf{r})\equiv h(\mathbf{r})
\end{equation}
in  Eq.~(\ref{eq:2.4}), which} guarantees $0<M(\mathbf{r})<1$ and has $M(\mathbf{r})\approx1$ for $r<r_0$, and $M(\mathbf{r})\approx 0$ for $r>r_0$ (i.e.\ \rp{the pattern is fixed outside of the sphere, while it is allowed to evolve within the sphere with mobility $M_0=1$). In these simulations, we monitor how the density evolves within the spherical region as it returns to a stationary amorphous structure.} As illustrated in Figs.~\ref{fig:3.21}(c) and \ref{fig:3.21}(d), a typical system evolves into a static bulk amorphous state for large times, but the configuration \rp{within the spherical region} may differ from that of the reference state. The difference between the \rp{system started from the initial condition described by Eq.~(\ref{eq:3.1})} and the reference state can be characterised by the following norm
\begin{equation}
\label{eq:3.4}
d(t) =\sqrt{\frac{1}{V}\int_\Omega dV \left[\phi(\mathbf{r}, t)-\phi^*(\mathbf{r})\right]^2},
\end{equation} 
which converges to zero if $\phi(\mathbf{r}, t)\rightarrow\phi^*(\mathbf{r})$. We use $d(t_f)\approx d(\infty)$, to measure the distance between the evolved structure and the reference system. We set $t_f = 3\times10^5$ to be the final time, which is long after the system ceases to measurably change in time. \rp{Note also that since the free energy density of the converged bulk amorphous solutions vary slightly as a function of the initial noise amplitude (see Fig.~\ref{fig:3.32}), all reference configurations were generated by using the same initial noise amplitude ($\chi=0.01\bar\psi$) to prevent the locally-melted system from systematically converging to a lower energy state. }

\begin{figure*}
 \includegraphics[width=0.99\linewidth]{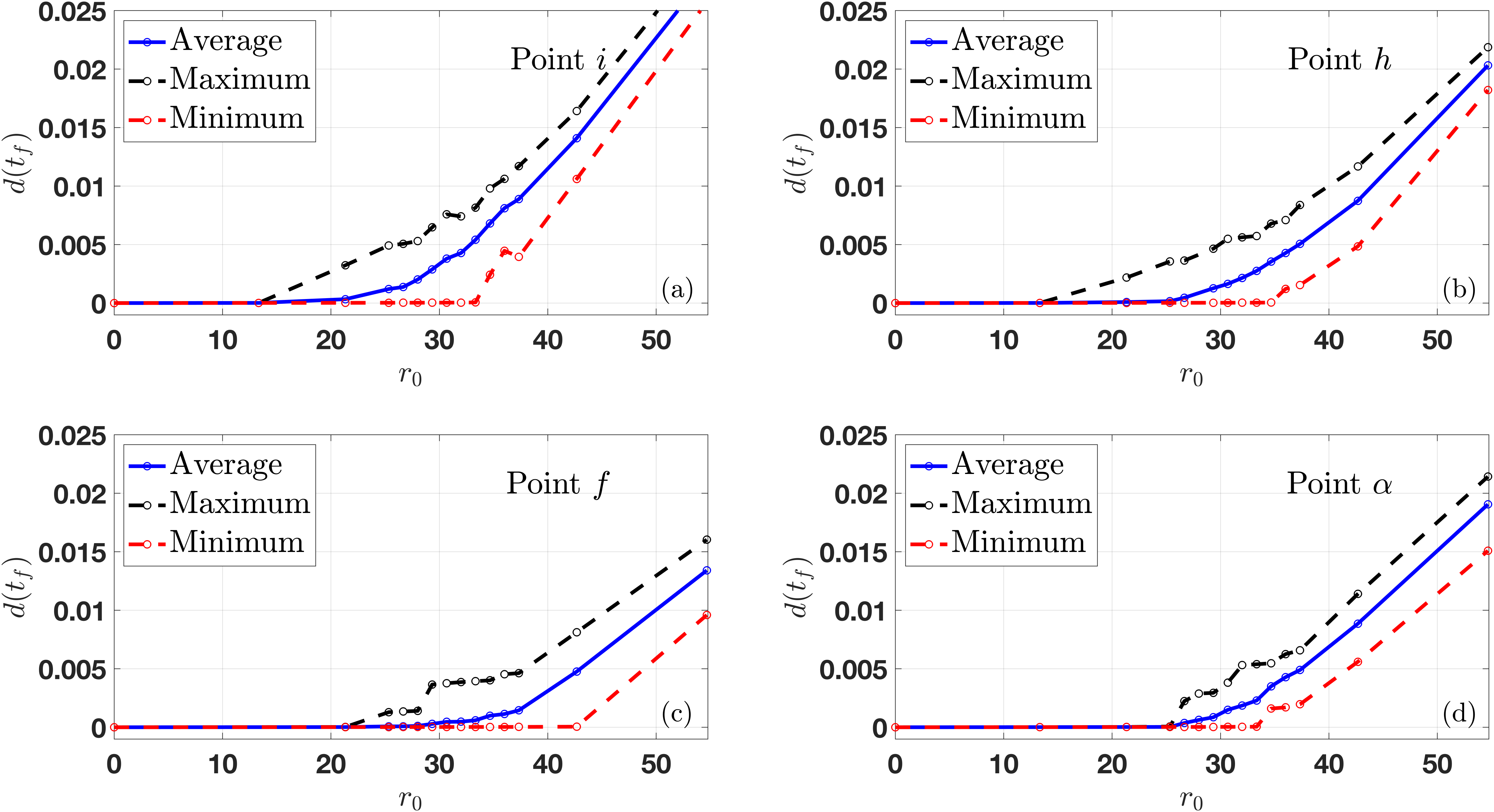}
  \caption{Plots of the norm $d(t_f)$ defined in Eq.~\eqref{eq:3.4}, as a function of melting radius $r_0$ [see Eqs.~\eqref{eq:3.1} and \eqref{eq:3.2}]. From this we can determine the point-to-set correlation length in the bulk amorphous state, defined as the largest value of $r_0$ for which we still have $d(t_f)\approx 0$. The results in panel (a) are for the state point $(\gamma, \bar\psi)=(0.275, -0.295)$, which is point $i$ in Fig.~\ref{fig:3.0}. In (b) $(\gamma, \bar\psi)=(0.197, -0.25)$, which is point $h$; in (c) $(\gamma, \bar\psi)=(0.146, -0.215)$, which is point $f$ and in (d) $(\gamma, \bar\psi)=(0.197, -0.255)$, which is point $\alpha$. The results are obtained for 28 different reference configurations. The maximum and minimum of $d(t_f)$ are indicated by the upper and lower dashed lines, while the solid curves show the average of $d(t_f)$ over the 28 configurations.}
 \label{fig:3.6}
 \end{figure*}
 
  \begin{figure}
 \includegraphics[width=0.99\linewidth]{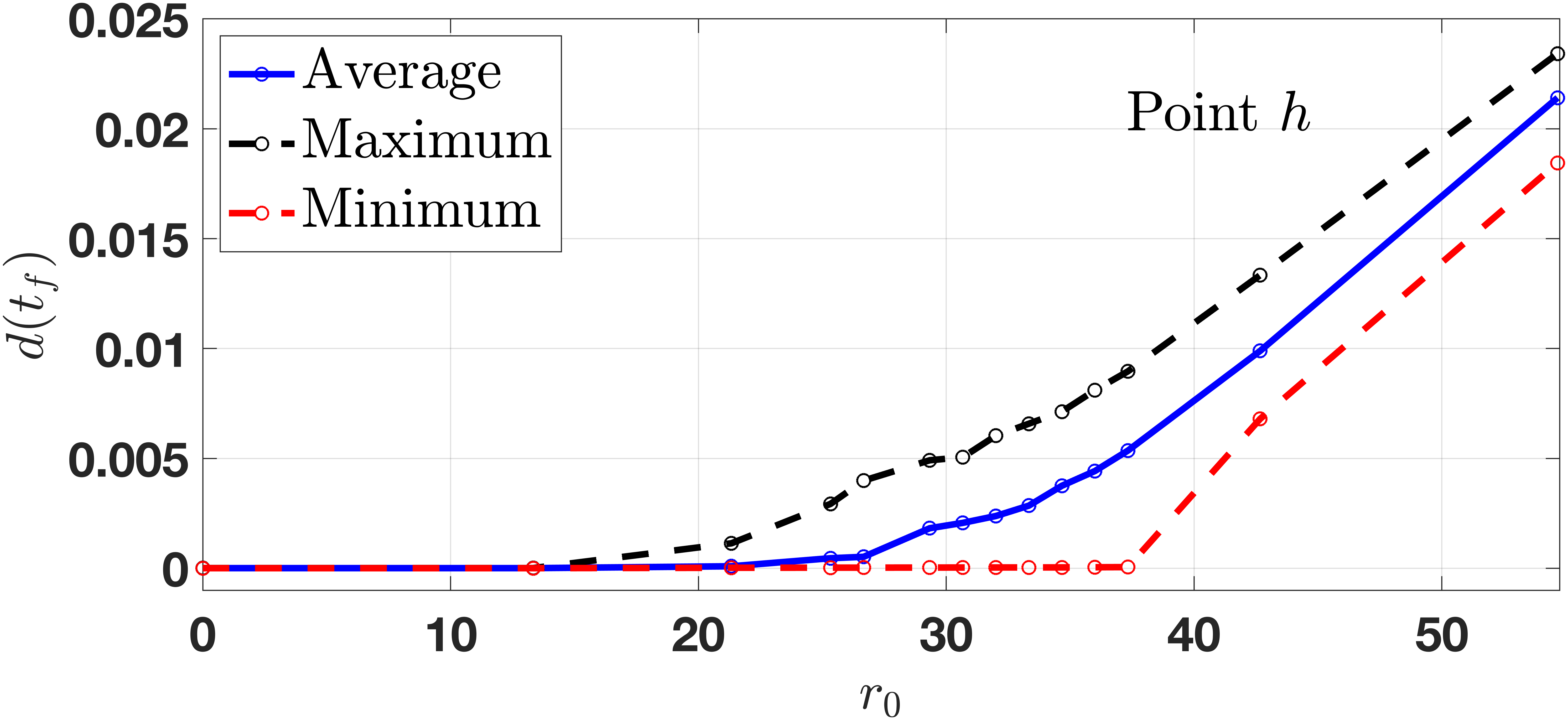}
   \caption{The norm $d(t_f)$ as a function of melting radius $r_0$, from which we can determine the point-to-set correlation length in the bulk amorphous state. These results are for point $h$ in the phase diagram, where $(\gamma, \bar\psi)=(0.197, -0.25)$ and initial noise amplitude $\chi=10\bar\psi$ (c.f.\ the results in Fig.~\ref{fig:3.6}(b), which are for $\chi=10^{-2}\bar\psi$).}
 \label{fig:3.5}
 \end{figure}

At each selected state point $(\gamma,\bar\psi)$ in the phase diagram, we repeated the simulation for 28 different reference configurations and for a range of 14 different values of the melting radius $r_0$. To \rp{understand how the point-to-set correlation length} $l_0$ depends on the location of the state point $(\gamma,\bar\psi)$, and in particular the distance from the critical point at $(\gamma,\bar\psi)=(0,0)$ and the distance from the liquid linear stability limit \rp{curve $\gamma=3\overline{\psi}^2$ (dashed line in Fig.~\ref{fig:3.0})}, the procedure was repeated at four different state points: the points $i$, $h$, $f$ and $\alpha$ on the phase diagram in Fig.~\ref{fig:3.0}. The results are summarised in Fig.~\ref{fig:3.6}. In Fig.~\ref{fig:3.6}(a), which is for state point $i$, the dependence of the norm $d(t_f)$ on \rp{the} radius $r_0$ shows that complete reconstruction of the reference amorphous state \rp{occurs for $r_0<14$ in all cases, while all final converged states differ from the reference state for $r_0>34$.} Intuitively, the point-to-set correlation length at fixed model parameters can be defined as the $r_0$ value at which the average of $d(t_f)$ (over many reference configurations) departs from 0. \rp{Alternatively, one can take the $r_0$ value at which exactly half of the configurations return to the reference configuration. For state point $i$, we find $l_0\approx 21$.} Recall that in the PFC model the typical distance between density maxima \rp{(also called atomic distance) is approximately $2\pi$. Thus, $l_0\approx21$ corresponds to a sphere of radius of $\approx 3.3$ atomic distances.}

Moving from point $i$ in the phase diagram towards the critical point at $(0,0)$ and going to points $h$ and $f$ we see from the results in Figs.~\ref{fig:3.6}(b) and \ref{fig:3.6}(c) that there is a modest increase in the value of $l_0$, becoming \rp{$l_0\approx25$ at point $h$ and $l_0\approx30$} at point $f$. Comparing Figs.~\ref{fig:3.6}(b) and \ref{fig:3.6}(d), which correspond to moving closer to the liquid stability limit (see point $\alpha$ in Fig.~\ref{fig:3.0}), we see no \rp{significant} change in the value of $l_0$.

To investigate the effect of the initial noise amplitude $\chi$ when generating the bulk reference states, we also measured the correlation length $l_0$ at state point $h$ for 32 configurations generated using the much larger initial noise amplitude $\chi=10\bar\psi$. These results are displayed in Fig.~\ref{fig:3.5}. Comparing these with the results in Fig.~\ref{fig:3.6}(b), which are for $\chi=10^{-2}\bar\psi$, we see no significant difference between the results obtained for these two very different values of the noise. In other words, the value of the point-to-set correlation length $l_0$ for amorphous structures in the PFC model does not depend on the history of how the states are prepared.

\rp{Considering the definition of the point-to-set correlation length $l_0$, our results show that one can model the amorphous structures in the PFC model as a mosaic of independent} regions of volume $V_0\approx \tfrac43\pi l_0^3$\rp{. Within} each of these volumes, the locations of the particles (density peaks) are determined by the locations of the others within that volume. In contrast, \rp{regions} separated by distances $\gg l_0$ are uncorrelated. Moreover, as \rp{discussed} already, we find that the free energy for systems of volume $V\gg V_0$, by the central limit theorem are normally distributed with variance $\sigma^2 =\sigma_0^2/n$, where $n=V/V_0$. Thus, the free energy \rp{density} may be written as
\begin{equation}
\label{eq:3.4_2}
f =\frac{1}{V}\int_\Omega dV f(\phi, \nabla\phi,...)=\frac{1}{n} \sum_{i=1}^n f_i,
\end{equation} 
where the free energy density of block $i$ on domain $\Omega_i$ is $f_i=\frac{1}{V_0}\int_{\Omega_i} dV f(\phi, \nabla\phi,....)$. Assuming that $f_i$ follows a certain random distribution with variance $\sigma_0$, it follows from the central limit theorem that the variance of $f$ depends on the system size as $\sigma^2 \propto 1/V$, which we verify numerically -- see Fig.~\ref{fig:3.3}(b). \rp{Note that this need not have been the case for the smallest of the simulation boxes that we considered. Since the volume of our simulation box is $(N\Delta x)^3$ and taking $l_0\approx30$, therefore the number of independent regions (blocks) in our smallest simulation volumes is $\approx[128(2/3)]^3/(\frac{4}{3}\pi 30^3)\approx5.5$, which of course is not large enough for the central limit theorem to apply. In contrast, for the largest simulation volumes the number of independent blocks is $\approx[1024(2/3)]^3/(\frac{4}{3}\pi 30^3)\approx2800$, for which we can be confident that the central limit theorem applies to determine the value of $f$.}

 \subsection{Stability of the amorphous \rp{structures}}
 \label{sec:3.3}
 
 \subsubsection{Unstable Liquid Region (ULR)}
 
\begin{figure}
 \includegraphics[width=0.99\linewidth]{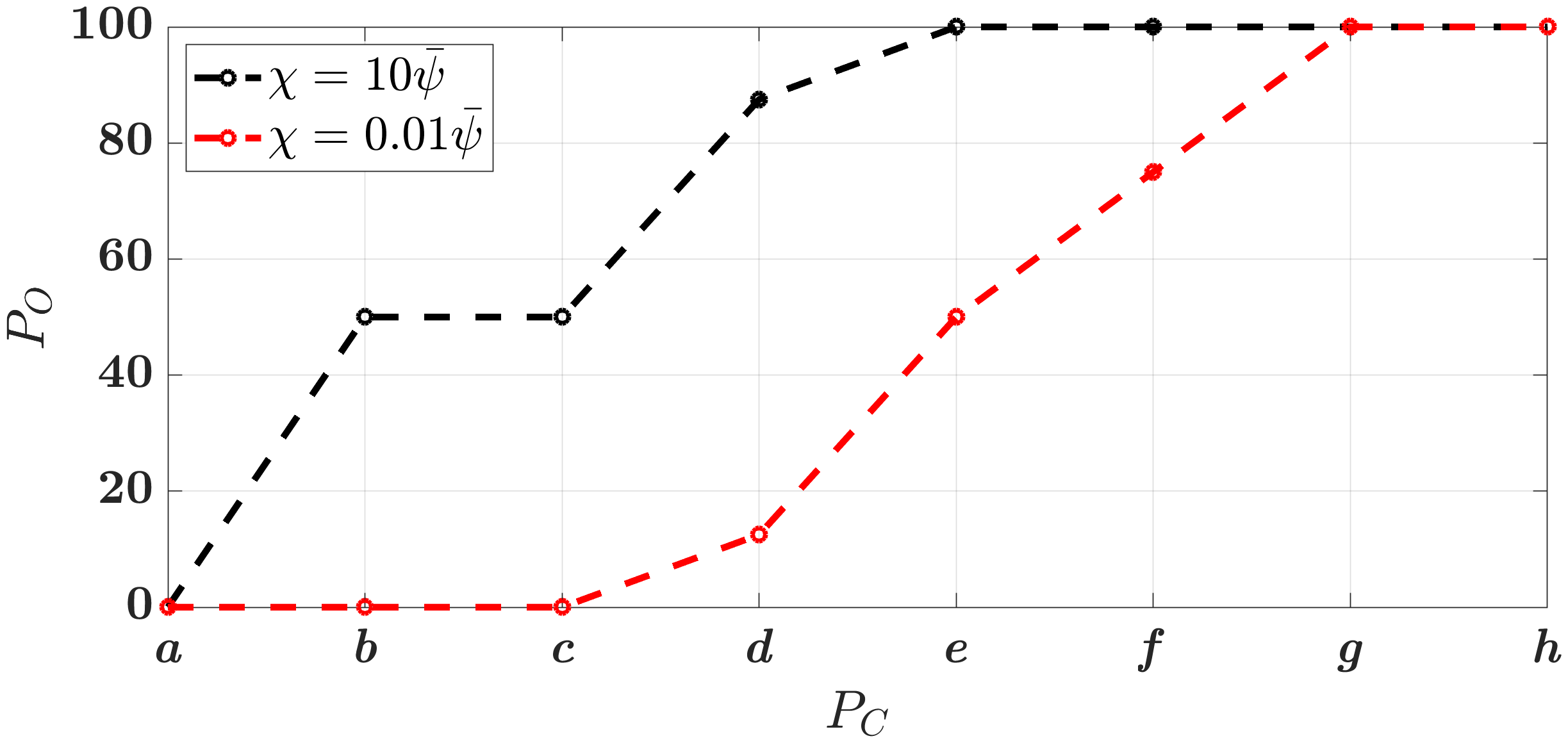}
  \caption{Amorphous-to-crystalline occurrence percentage $P_O$ for two different values of the initial noise amplitude, $\chi= 10^{-2}\bar\psi$ and $\chi= 10\bar\psi$, at the sequence of state points $P_C=\{a,b,c,...,h\}$ in the phase diagram in Fig.~\ref{fig:3.0}. The results are for $N=512$ and averaging over 8 simulations at each state point.}
 \label{fig:3.8}
 \end{figure}

We now discuss the stability of the bulk amorphous state at different locations on the PFC phase diagram. In the unstable liquid regime (ULR), which is to the right of the dashed linear-stability threshold line in the phase diagram in Fig.~\ref{fig:3.0}, the homogeneous liquid state is unstable, with all density perturbations leading to a spontaneous barrierless transition to either the crystal or the amorphous state. Numerical solutions of Eq.~(\ref{eq:2.4}) indicate that far from the critical point, typically for $\gamma>0.19$, such as at state points $\alpha$, $h$ and $i$ in Fig.~\ref{fig:3.0}, stable amorphous configurations grow from the melt for all initial noise amplitudes $\chi>0$ (we tested in the range $10^{-7}\bar\psi<\chi<10\bar\psi$). However, on moving \rp{closer to the critical point ($\gamma<0.12$), and for densities roughly in the vicinity of the liquid stability line $\gamma=3\bar{\psi}^3$,} the BCC crystalline state becomes the final state formed after quenching \rp{for any} $\chi>0$ (we checked in the range $10^{-7}\bar\psi<\chi<23\bar\psi$). This indicates \rp{that} the amorphous state becomes unstable for $\gamma<0.12$. To characterise the transition from crystal-formation to glass-formation on quenching into the ULR, we performed a series of simulations (8 different configurations at each point) with noise amplitudes $\chi_1=0.01\bar\psi$ and $\chi_2=10\bar\psi$ at points $a$--$h$ in the phase diagram, Fig.~\ref{fig:3.0}. These state points lie parallel to the liquid stability threshold and to the right of it. These results are summarised in Fig.~\ref{fig:3.8}, which shows the percentage of the simulations at each state point that ended up in an amorphous state, as opposed to the BCC crystal. We see that solely bulk crystalline states are formed at state point $a$, while just bulk amorphous states are obtained at state point $g$. However, both states can arise as the final state in a relatively broad range in between these two end points. This amorphous-to-crystal ratio of the final states increases with increasing distance from the critical point, which indicates a smooth transition regime, rather than a sharp amorphous stability line (such as the liquid stability threshold). The amorphous-to-crystalline ratio also depends on the initial noise. Figure \ref{fig:3.8} shows that the transition to the amorphous state can occur closer to the critical point and with higher probability if the initial noise amplitude is larger. This indicates that strong initial perturbations prevent the formation of the crystal in the ULR.

 \begin{figure}
 \includegraphics[width=0.99\linewidth]{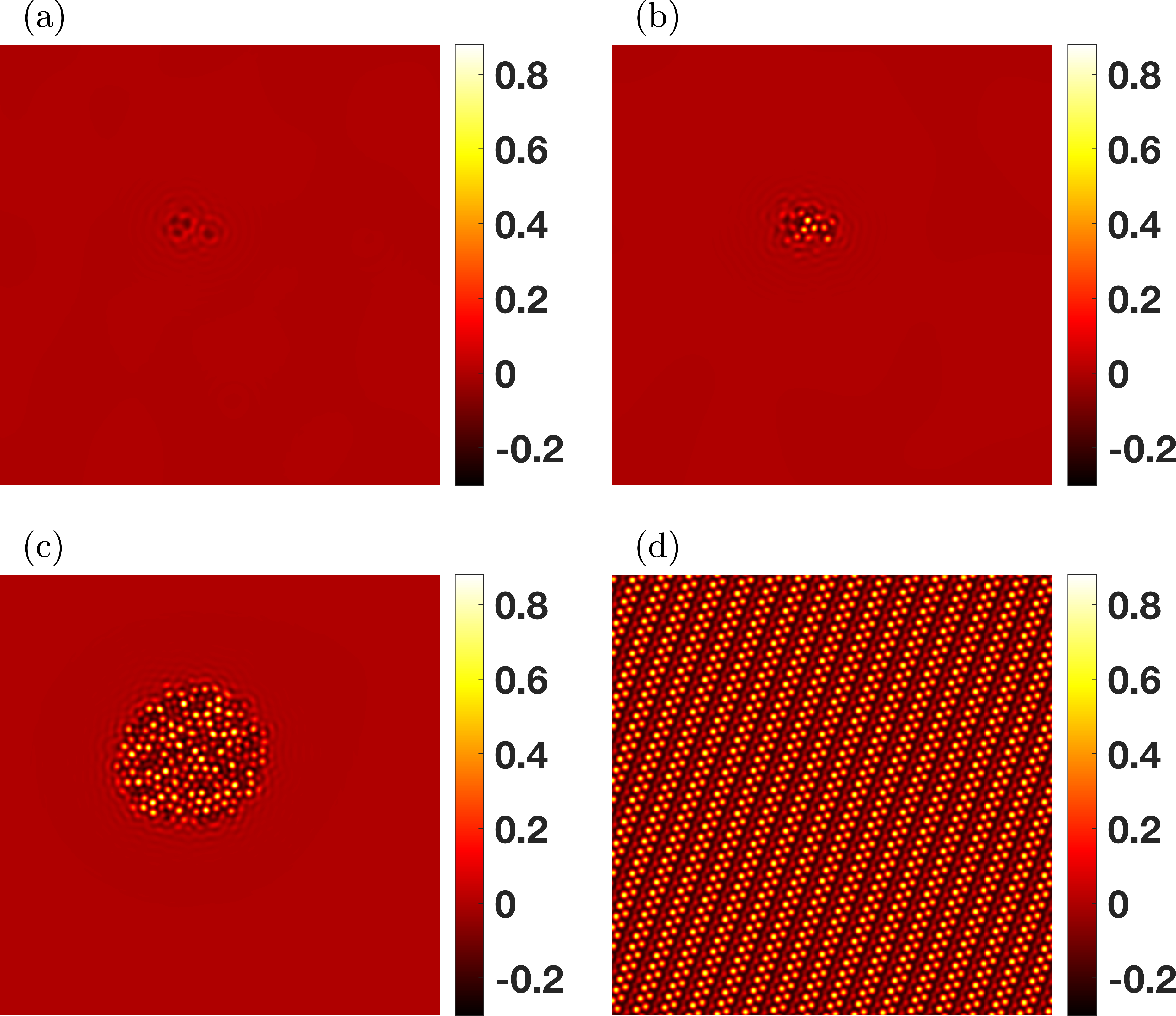}
    \caption{Formation of the BCC crystal by first nucleating an amorphous cluster that subsequently rearranges into the crystal. We display a two dimensional cross section of a $N=512$ size system at the state point $(\gamma, \bar\psi)=(0.189, -0.281)$ (point $\beta$ in Fig.~\ref{fig:3.0}) with initial noise amplitude $\chi=10\bar\psi$. These simulation snapshots are at the times (a) $t=375$, (b) $t=750$, (c) $t=2625$, and (d) $t=125000$.}
    \label{fig:3.212}
\end{figure}

 \begin{figure}
 \includegraphics[width=0.99\linewidth]{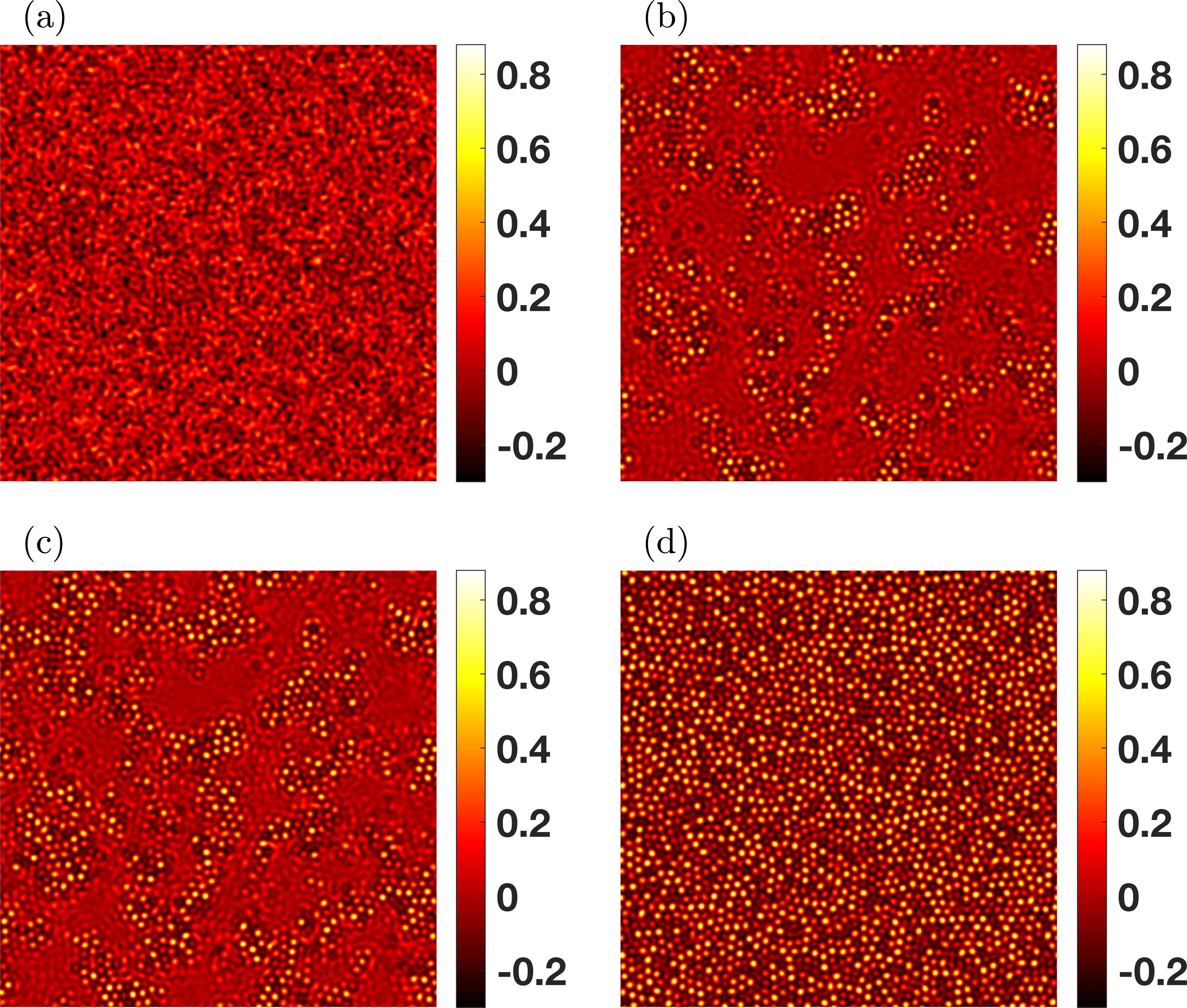}
    \caption{Formation of the amorphous from multiple nucleation sites. We display a two dimensional cross section of a $N=512$ system at the state point $(\gamma, \bar\psi)=(0.189, -0.265)$  with initial noise amplitude $\chi=10\bar\psi$. The simulation snapshots are at the times (a) $t=0$, (b) $t=125$, (c) $t=150$, and (d) $t=125000$.}
    \label{fig:3.209}
\end{figure}

  \begin{figure}
 \includegraphics[width=0.99\linewidth]{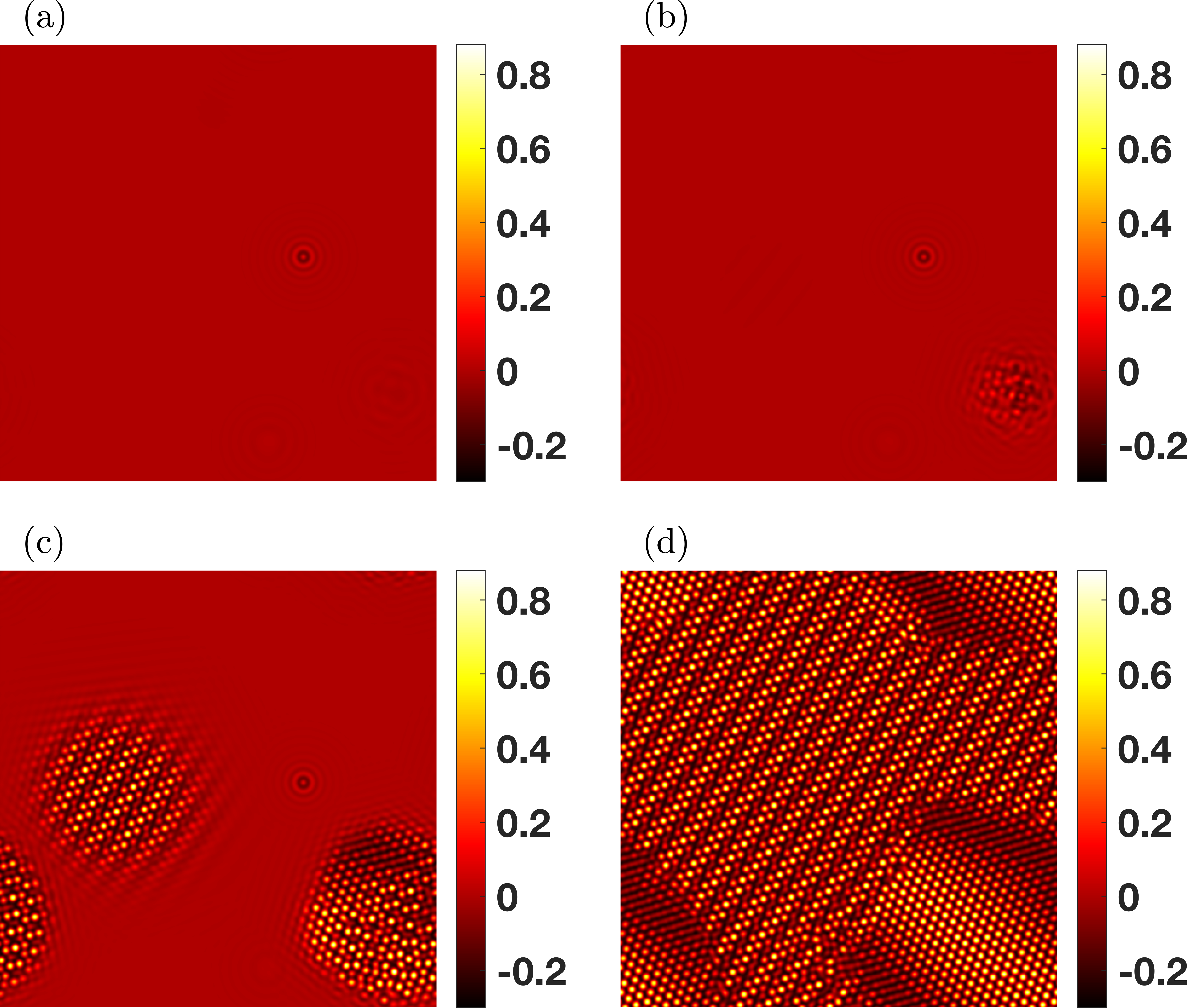}
    \caption{Results from state point $n$, the same as in Fig.~\ref{fig:3.209}, except here the initial noise amplitude is the smaller value $\chi=5\bar\psi$. Here we observe the formation of several nuclei, all with crystalline ordering. The simulation snapshots are at the times (a) $t=875$, (b) $t=1125$, (c) $t=150$, and (d) $t=125000$.} 
    \label{fig:3.210}
\end{figure}

  \begin{figure}
 \includegraphics[width=0.99\linewidth]{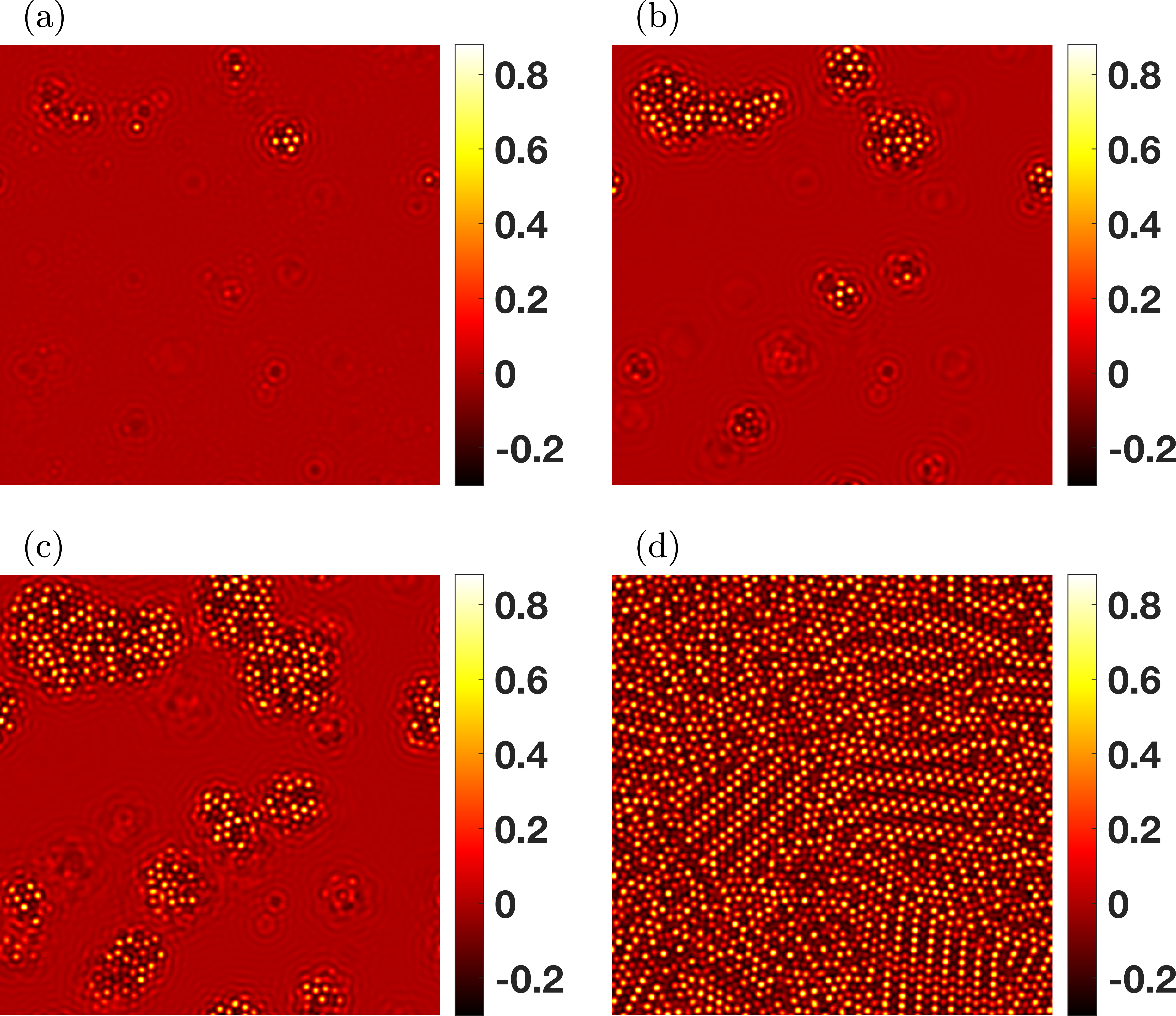}
    \caption{Results from state point $n$, the same as in Fig.~\ref{fig:3.209}, except here the initial noise amplitude is $\chi=6\bar\psi$. For this intermediate value of the noise we observe the nucleation of both the amorphous and crystalline structures, leading to a final state that is a mixture of both. The simulation snapshots are at the times (a)$t=175$, (b) $t=350$, (c) $t=525$, and (d) $t=175000$.} 
    \label{fig:3.211}
\end{figure}
 
 \subsubsection{Stable Liquid Region (SLR)}
 
The most significant difference between the ULR and the stable liquid region (SLR), the former being to the right of the dashed line in the phase diagram in Fig.~\ref{fig:3.0}, is that the liquid is metastable in the SLR. Consequently, the phase transition from the liquid phase is not barrierless, i.e.\ a big enough perturbation amplitude $\chi$ is required to induce a phase transformation from the melt. We have verified this numerically. For instance, at point $j$ in the phase diagram (see Fig.~\ref{fig:3.0}), the liquid phase is found to be stable for $\chi \leq 8 \bar\psi$, while $\chi > 8 \bar\psi$ induces a transition to the crystalline phase. Similarly to in the ULR, the amorphous state appears far from the critical point, \rp{and occurs with a probability increasing with the distance from the critical point.}

Since the phase transition process involves induced nucleation, it is worthwhile to investigate how the final structure depends on the distance from the liquid stability line at fixed $\gamma$. At state point $\beta$ in Fig.~\ref{fig:3.0}, which is close to the crystal-liquid coexistence regime, the nucleation barriers for forming both the crystal and amorphous phases are high. The formation of a non-crystalline solid seed (precursor) was observed first for initial noise amplitude $\chi=10\bar\psi$, and is shown in Fig.~\ref{fig:3.212}. This process is followed by a secondary transition, during which the growing solid seed undergoes a structural rearrangement, resulting in a bulk crystalline final structure -- see the bottom right panel of Fig.~\ref{fig:3.212}.

Moving in the phase diagram closer to the liquid stability limit at constant $\gamma$ (see point $n$ in Fig.~\ref{fig:3.0}), the nucleation barrier decreases due to the increasing supersaturation. Under these circumstances, both the nucleation process and the final state depended on the initial noise amplitude. For the higher noise amplitude value of $\chi=10\bar\psi$, we observe the formation of many nucleation centres which quickly interlock and the final state is amorphous -- see Fig.~\ref{fig:3.209}. In contrast, for the smaller noise amplitude value of $\chi=5\bar\psi$, the nuclei that form display crystalline order, and we only observe a crystalline final state -- see Fig.~\ref{fig:3.210}. Between these two extremes, \rp{e.g.\ at} $\chi=6\bar\psi$, competing nucleation and growth of both amorphous and crystalline seeds are observed, leading to a final state that is a mixture of both -- see Fig.~\ref{fig:3.211}.

  \subsection{Coexistence of crystal and amorphous}
  \label{sec:3.4}

\begin{figure*}
 \includegraphics[width=0.99\linewidth]{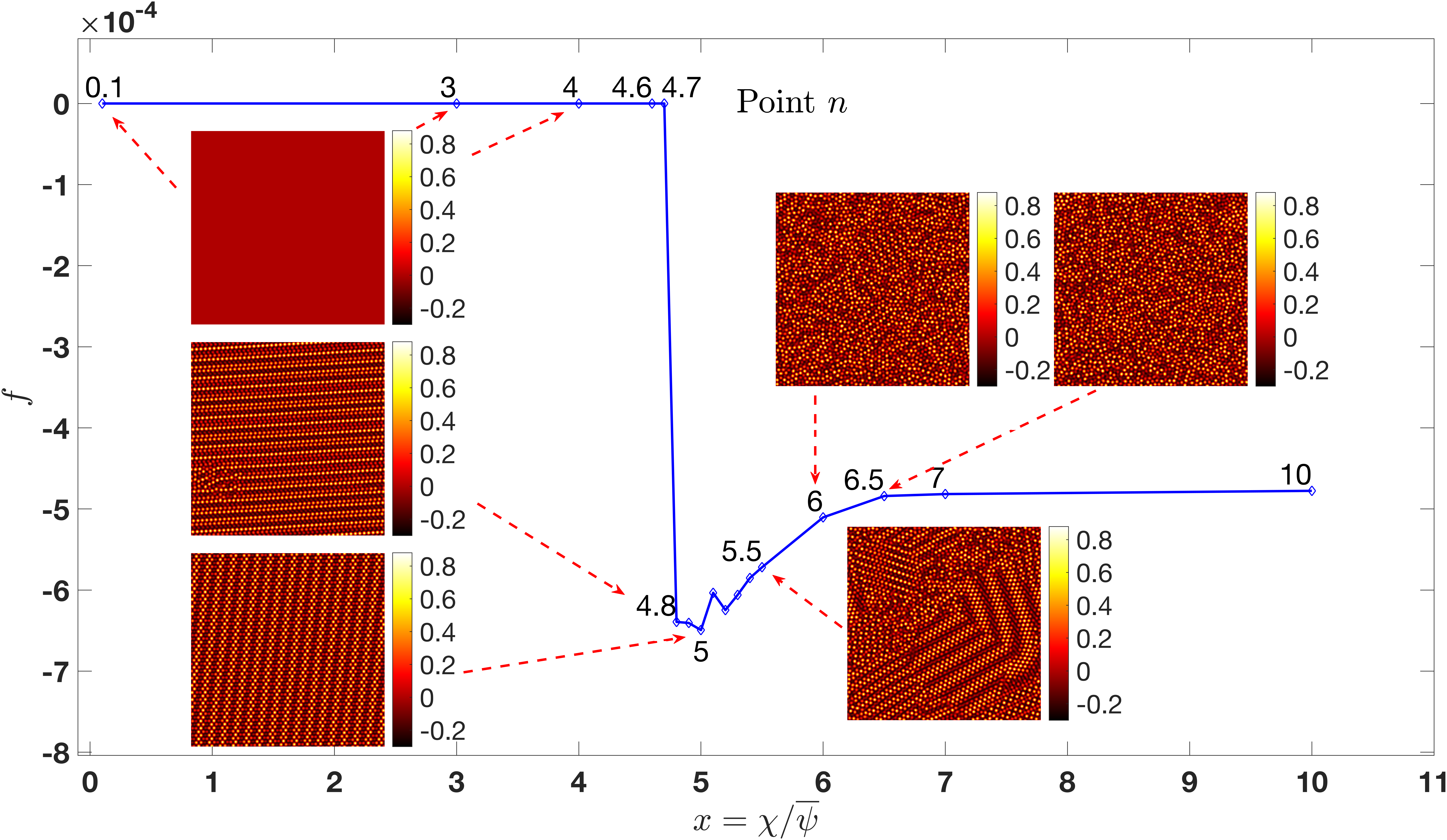}
  \caption{The free energy density $f=F/V$ of the final state as a function of $x=\chi/\bar\psi$, where $\chi$ is the amplitude of the noise in the initial profile, at state point $n$ in Fig.~\ref{fig:3}, where $(\gamma, \bar\psi)=(0.189, -0.265)$. The various insets display slices through corresponding density profiles for the $x$ values indicated.}
 \label{fig:3.10}
 \end{figure*}

To elucidate further the effect of the initial noise strength $\chi$ on the final state of the system at state point $n$ in the phase diagram, we have performed simulations over a broad range of values of $\chi$. The results are summarised in Fig.~\ref{fig:3.10}, which displays a plot of the free energy density $f=F/V$ as a function of $\chi$ as well as typical plots of slices through the final density profiles. For $\chi \lesssim 4.7\bar\psi$ the initial perturbations decay, and the system relaxes to the metastable liquid state. For $\chi \approx 4.8\bar{\psi}$, the initial condition locally surmounts the nucleation barrier for direct crystal nucleation at a few locations, thus leading to the formation of the bulk crystalline phase. With increasing initial noise amplitude, more and more nucleation centres appear, from which crystalline grains of different orientations grow. When the grains meet, grain boundaries form, and therefore the final state is polycrystalline -- see e.g.\ the density profile cross section corresponding to $\chi=5.5\bar{\psi}$ in Fig.~\ref{fig:3.10}. Since the grain boundaries lead to a positive contribution to the total free energy $F$, the value of $f=F/V$ for these configurations is higher than for those where the system forms a single crystal grain, having a contribution to $F$ that is proportional to the total grain boundary area in the system. Interestingly, increasing the initial noise amplitude further into the range $5.5\bar{\psi}\lesssim\chi\lesssim6\bar{\psi}$ does not lead to the formation of a nano-crystalline final state. Instead, competing nucleation and growth of crystalline and amorphous grains is observed due to the fact that the large-amplitude initial perturbations locally pass the (larger) nucleation barrier for direct amorphous nucleation. One can see in Fig.~\ref{fig:3.10} in the cross section corresponding to $\chi=6\bar{\psi}$ that the final state is a mixture of coexisting amorphous and crystalline domains with crystalline-amorphous grain boundaries. Similar final states were also found at points $k$, $l$, $m$, and $o$ of the phase diagram in Fig.~\ref{fig:3.0}. Finally, for $\chi\geq6.5\bar{\psi}$, crystalline grains are no longer detectable, and the final state is bulk amorphous.  
 
\section{Summary}
\label{sec:4}

In this study, based on numerical simulations of the PFC model, which is a simplified classical DFT for condensed matter, we have determined the thermodynamic properties and the nature of the microscopic structural correlations of the amorphous phase that forms from the undercooled liquid via rapid solidification processes. We have evaluated the distribution of free energy density values $f$ of metastable bulk amorphous configurations that form in the region of the phase diagram where the BCC crystal is the global free energy minimum. For systems that are all prepared in the same manner, we have found that the average free energy density $\bar{f}$ converges to a well defined value, i.e.\ it has a well-defined value in the thermodynamic limit. However, we have also found that the value of $\bar{f}$ does depend on the formation dynamics, in particular systems that form from initial states with (random) density fluctuations that have a large amplitude $\chi$ have a slightly larger value of $\bar{f}$ than those that form from states with a smaller value of $\chi$. Roughly speaking, large values of $\chi$ correspond to a deep quench in temperature from the molten liquid to form the amorphous solid, while smaller values of $\chi$ correspond to a more shallow quench to form the amorphous.

These results indicate that one may view the amorphous state as being made up of a patchwork (mosaic) of regions where the free energy density varies from patch to patch and its value is not correlated (or only weakly correlated) to the value in neighbouring patches. This picture is supported by our measurements of the point-to-set correlation length $l_0$ \cite{berthier2011theoretical, bouchaud2004adam}, which is defined as the limit radius for which the original amorphous pattern reforms unchanged after a spherical volume of the amorphous solid is locally melted and then allowed to re-solidify. Beyond this limit radius, spatial points can be regarded as uncorrelated. Our results show that the correlation length $l_0$ is finite, with only modest variations in value on moving to different state-points in the the phase diagram. Since the bulk amorphous consists of independent domains with a certain free energy distribution and characteristic volume $V_0\approx\tfrac43\pi l_0^3$, the central limit theorem of course implies a well defined value for the free energy density in the infinite volume limit.

We have also studied the stability properties of the amorphous configurations. We found that the stability of the amorphous phase depends greatly on the distance in the phase diagram from the critical point on both sides of the liquid stability curve. We find that the amorphous phase does not form for $\gamma < 0.12$ (i.e.\ it is unstable in this regime), with the likelihood of forming increasing as $\gamma$ is increased and also as the amplitude of the noise in the initial state $\chi$ is increased.

In the final part of our \rp{investigation}, we studied the microscopic aspects of the formation dynamics of the amorphous configurations. In particular, we monitored the forming solid structures as a function of the model parameters and the amplitude of the noise added to the homogeneous initial condition. Depending on the values of the parameters, we observed (i) direct amorphous nucleation with a subsequent barrierless (i.e. without the addition of any further noise) transition into the crystalline phase, (ii) direct amorphous nucleation leading to the bulk amorphous phase, (iii) direct crystal nucleation with a bulk crystalline final state, and (iv) competing nucleation of the crystalline and the amorphous phase, leading to a mixture of coexisting amorphous and crystalline grains. This variety of solidification pathways is in a good agreement with previous theoretical predictions \cite{Lutsko_PRL_2006, Berry_PRE_2008, PhysRevLett.107.175702, PODMANICZKY201724, TANG_2017, GRANASY_2019,Lutsko_SA_2019, Podmaniczky_2021} and experimental observations \cite{Schoepe_PRL_2006,Zhang_JACS_2007,Zhang_JPCB_2007,TanXuXu2014} of the so-called amorphous precursor mediated multi-step crystallisation process and the formation of a glassy phase (called the q-glass \cite{q_glass}) in a first order phase transition \cite{q_glass,l-g_coex}. 

We believe that our results contribute to the better theoretical understanding of the thermodynamic and structural properties of amorphous solids as well as their role in complex solidification processes. Besides, the results give further evidence for the qualitative applicability of the PFC model in studying general structural aspects of first order phase transitions involving solid structures. Even though the PFC model is far from representing real systems quantitatively, relating the model parameters to measurable physical quantities does help to predict the outcome of (and therefore can offer a way to control) solidification processes.

\section*{Acknowledgements}

AJA was supported by the EPSRC under the grant EP/P015689/1 and LG was supported by the National Agency for Research, Development, and Innovation (NKFIH), Hungary under contract No. KKP-126749.


%

\end{document}